\documentclass[english]{article}
\usepackage[T1]{fontenc}
\usepackage[latin9]{inputenc}
\usepackage{babel}
\usepackage{amsmath}
\usepackage{amssymb}
\usepackage{graphicx}
\usepackage{wasysym}
\usepackage[unicode=true,pdfusetitle,
 bookmarks=true,bookmarksnumbered=false,bookmarksopen=false,
 breaklinks=false,pdfborder={0 0 1},backref=section,colorlinks=false]
 {hyperref}

\makeatletter

\providecommand{\tabularnewline}{\\}

\newenvironment{lyxcode}
{\par\begin{list}{}{
\setlength{\rightmargin}{\leftmargin}
\setlength{\listparindent}{0pt}
\raggedright
\setlength{\itemsep}{0pt}
\setlength{\parsep}{0pt}
\normalfont\ttfamily}%
 \item[]}
{\end{list}}

\usepackage{authblk}
\usepackage{babel}

\makeatother

\begin{document}
\title{Gamra: Simple Meshing for Complex Earthquakes}

\author[1]{Walter Landry\thanks{wlandry@caltech.edu}}

\author[2]{Sylvain Barbot\thanks{sbarbot@ntu.edu.sg}}

\affil[1]{Infrared Processing and Analysis Center, Caltech, Pasadena, CA 91125, USA}

\affil[2]{Earth Observatory of Singapore, 50 Nanyang Avenue, Nanyang Technological University, 639798, Singapore}

\maketitle
\begin{abstract}
The static offsets caused by earthquakes are well described by elastostatic
models with a discontinuity in the displacement along the fault. A
traditional approach to model this discontinuity is to align the numerical
mesh with the fault and solve the equations using finite elements.
However, this distorted mesh can be difficult to generate and update.
We present a new numerical method, inspired by the Immersed Interface
Method \cite{leveque&li94}, for solving the elastostatic equations
with embedded discontinuities. This method has been carefully designed
so that it can be used on parallel machines on an adapted finite difference
grid. We have implemented this method in Gamra, a new code for earth
modelling. We demonstrate the correctness of the method with analytic
tests, and we demonstrate its practical performance by solving a realistic
earthquake model to extremely high precision. 
\end{abstract}

\section{Motivation}

A common feature of many earthquakes is a complex network of intersecting
faults. Accurately modeling the static offsets and associated large
scale deformation due to this fault geometry is crucial to a reliable
understanding of seismic hazards \cite{marshall+08}. The behavior
of these faults is relatively well described by the equations of variable
modulus elastostatics. However, for realistic faults, the displacement
does not gradually taper off, but rather ends abruptly. This abrupt
termination gives rise to a logarithmic singularity in the displacement
\cite{okada92}. In realistic faults, these singularities are smoothed
out by non-linear processes at the fault tips that are on a scale
that are many orders of magnitude smaller than the fault itself. These
characteristics make it challenging to numerically model realistic
fault networks.

In addition, elastostatics is only one piece of the puzzle when modeling
the earthquake cycle. We want to incorporate an elastostatic solver
into an overall algorithm for modeling the entire earthquake cycle
\cite{barbot+12}. We desire a unified method, using the same mesh,
architecture, and boundaries, that can solve elliptic equations (for
static offsets of earthquakes), parabolic equations (for poro-elastic
and visco-elastic evolution between earthquakes), and hyperbolic equations
(for dynamic rupture during an earthquake). Then we will have a powerful
tool for self consistent models of the entire earthquake cycle.

At present, one relatively successful approach to building this kind
of tool uses boundary integral methods \cite{barbot+12,kaneko+10,lapusta&barbot12,hori+04,kato04,matsuzawa+10,rice93,shibazaki&shimamoto07,smith&sandwell04a}.
However, boundary integral methods inevitably make simplifications
in the geometry or the physics of the problem. Finite-element methods
\cite{aagaard+13,hassani+97,melosh&williams89,puente+09,kaneko+08,kaneko+11}
provide a natural way to fully represent the geometry and the physics
as long as the mesh conforms to the faults. Generating these conforming
meshes can be quite challenging and time consuming, especially when
the faults intersect. The extended finite element method \cite{becker2009nitsche,coon+11,zangmeister2015extended}
shows great promise in addressing this problem with mesh generation,
though it has yet to be applied to realistic 3D earthquake models.

Finite difference methods, on the other hand, have not traditionally
been used for this kind of problem. Straightforward implementations
of finite differences require that the displacement be continuous
and differentiable. This limitation spurred the development of the
Immersed Interface Method (IIM) \cite{leveque&li94}. IIM explicitly
models the discontinuous jump, resulting in a series of corrections
to the ordinary finite difference stencils. IIM has spawned a number
of variations, and some of these have been applied to various problems
in elastostatics \cite{rutka2006ejiim,rutka2006explicit,botella2010ls,zhu2012second}.
None of them have looked at models most relevant to earthquakes, where
we prescribe the discontinuity in the displacement. More importantly,
none of them have discussed how to handle the difficulties associated
with the singularity at the fault tip. Finally, none of these methods
have been implemented on adapted grids or parallel machines.

The purpose of this paper is to describe a new method, inspired by
IIM, that naturally handles all of the difficulties associated with
faults. This method was developed with an eye towards performance,
so it naturally extends to the use of parallel machines and highly
adapted grids. With this solver in place, we can then use the existing
deep understanding of how to implement hyperbolic and parabolic solvers
for the equations specific to earthquakes in a finite difference framework
\cite{day82,dunham&archuleta05,dunham+11,andrews02a,day+05,harris+09,olsen+97,ely+09,ely+10,cui+10,kozdon+13,moczo+14}.

We first describe the equations of linear elasticity, how we treat
internal dislocations, and how we solve these equations on an adapted
mesh. Then we demonstrate the correctness of the method and our implementation
with a series of analytic tests. Finally, we document the performance
of our implementation with a simulation of the 1992 Mw 7.3 Landers
earthquake. The algorithm described in this paper is implemented in
Gamra, a code available at \href{https://bitbucket.org/wlandry/gamra}{https://bitbucket.org/wlandry/gamra}.
Gamra is a French acronym for Géodynamique Avec Maille Rafinée Adaptivement,
meaning ``geodynamics with adaptive mesh refinement''.

\section{Methods}

We begin by describing the equations of linear elasticity (section
\ref{subsec:Governing-Equations}) and the mesh we use for solving
them (section \ref{subsec:Staggered-Grid}). Then we describe the
Gauss-Seidel smoother that we use as a component in our solvers (section
\ref{subsec:Gauss-Seidel-Relaxation}). Then we describe the corrections
we make to treat internal dislocations of arbitrary orientation in
two and three dimensions (section \ref{subsec:Fault-Corrections}).
Then we describe how we implement boundary conditions (section \ref{subsec:Boundary-Conditions}).
With these components, we have a stable, accurate solver for earthquake
physics.

However, this will not be a fast solver without multigrid. To implement
multigrid (section \ref{subsec:Multigrid}), we need coarsening (section
\ref{subsec:Coarsening}) and refinement (section \ref{subsec:Refinement})
operators. To implement adaptive multigrid, we also need to set boundary
conditions at coarse-fine boundaries (section \ref{subsec:Coarse-Fine-boundaries}).

\subsection{Governing Equations\label{subsec:Governing-Equations}}

We solve the Navier's equation for elastostatic deformation with the
infinitesimal strain approximation 
\begin{equation}
\sigma_{ji,j}+f_{i}=0~,\label{eq:elasto-statics}
\end{equation}
where the stress components $\sigma_{ji}$ are defined using Hooke's
law in terms of the displacement components $v_{i}$, Lame's first
parameter $\lambda$, and the shear modulus $\mu$ 
\begin{equation}
\sigma_{ji}\left(\vec{v}\right)\equiv\mu(v_{i,j}+v_{j,i})+\delta_{ij}\lambda v_{k,k}.\label{eq:Stress}
\end{equation}
We use Einstein summation notation, where each index $i$, $j$, $k$
is understood to $x$, $y$, and $z$ in turn, repeated indices are
summed, and commas (,) denote derivatives.

For all of our test problems, the stress tensor will be symmetric
$\left(\sigma_{ij}=\sigma_{ji}\right)$. In addition, the forcing
term $f_{i}$ is zero for many of our test problems. But equivalent
body forces can be used represent inelastic deformation in quasi-static
deformation simulations \cite{barbot&fialko10b,rousset+12,rollins+15}.
Therefore the inclusion of body forces in Eq. (\ref{eq:elasto-statics})
is critical for modeling quasi-static deformation due to off-fault
processes.

\subsection{Staggered Grid\label{subsec:Staggered-Grid}}

\begin{figure}
\begin{centering}
\includegraphics[width=0.25\columnwidth]{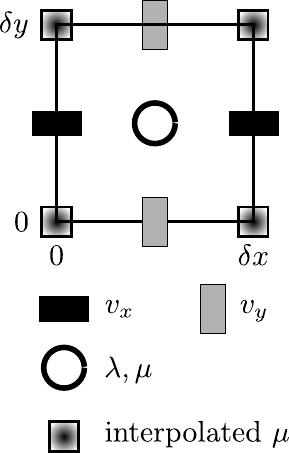}
\par\end{centering}
\caption{Reference cell showing where the displacement and moduli are defined.
The bottom left is at $x=0$, $y=0$, and the top right is at $x=\delta x$,
$y=\delta y$.\label{fig:Reference-cell}}

\end{figure}

We discretize the equations on a staggered grid, with the displacement
located at cell faces as shown in Figure \ref{fig:Reference-cell}.
Our method requires the shear modulus ($\mu$) at both the cell centers
and cell corners. Since $\mu$ is a given function of space, we could
compute it exactly at both cell centers and corners. We have found
that we get larger reductions in the residuals for each multigrid
V-cycle by using the given function to compute the cell centers, and
then using the geometric mean to fill the value at the cell corners.
Specifically, in 2D, for a reference cell where the bottom left corner
is located at $x=0$, $y=0$, $\mu$ at that corner is 
\begin{equation}
\left.\mu\right|_{0,0}=\left(\left.\mu\right|_{\delta x/2,\delta y/2}\left.\mu\right|_{-\delta x/2,\delta y/2}\left.\mu\right|_{\delta x/2,-\delta y/2}\left.\mu\right|_{-\delta x/2,-\delta y/2}\right)^{1/4}.\label{eq:modulus_geometric_average}
\end{equation}
The subscripts $\left.\right|_{0,\delta y/2}$ indicate the variable
located at an offset of $x=0$, $y=\delta y/2$ from the bottom left
corner. So $\left.\right|_{0,0}$ is the bottom left corner, $\left.\right|_{0,\delta y/2}$
is the left face, and $\left.\right|_{\delta x/2,\delta y/2}$ is
the cell center.

The Lame parameter $\lambda$ is only needed at cell centers, so there
is no extra interpolation step. 

We can specify $\mu$ and $\lambda$ one of two ways: analytic expressions
and tables. We use the muparser library \cite{muparser-home-page}
to evaluate analytic expressions. To compute the modulus at the boundary,
we may need the modulus at a point outside the boundary. For analytic
expressions, we evaluate the expression at that outside point. For
moduli given by a table, we choose the closest point covered by the
table.

For multigrid, the modulus on coarser levels is interpolated from
finer levels, not directly computed. Using the interpolated values
rather than the directly computed values results in larger reductions
in the residuals for each multigrid V-cycle. The interpolation onto
the cell centered modulus is a simple arithmetic average of all of
the fine points in the coarse cell.

This treatment of the modulus works well for the moderate jumps in
material properties seen in realistic models of earthquake regions.
More extreme jumps would require a more sophisticated treatment, such
as applying IIM to material interfaces as well as faults.

\subsection{Gauss-Seidel Relaxation\label{subsec:Gauss-Seidel-Relaxation}}

The core of the solver is a red-black Gauss-Seidel relaxation. We
first define the residual as the non-zero remnant of equation \ref{eq:elasto-statics} 

\begin{equation}
r_{i}\left(\vec{v},\vec{f}\right)=\sigma_{ji,j}+f_{i}~.\label{eq:residual}
\end{equation}
We discretize the residual in the usual way with centered differences.
To be explicit, in 2D, we write the $x$ component as 
\[
\sigma_{jx,j}=\left(\left(\lambda+2\mu\right)v_{x,x}\right)_{,x}+\left(\lambda\,v_{y,y}\right)_{,x}+\left(\mu\left(v_{x,y}+v_{y,x}\right)\right)_{,y}~.
\]
where, in the reference cell 
\begin{equation}
\begin{aligned}\big(v_{x,x}\big(\lambda & +2\mu\big)\big)_{,x}\bigg|_{0,\delta y/2}=\\
 & \left[\left(\left.v_{x}\right|_{\delta x,\delta y/2}-\left.v_{x}\right|_{0,\delta y/2}\right)\left(\left.\lambda\right|_{\delta x/2,\delta y/2}+2\left.\mu\right|_{\delta x/2,\delta y/2}\right)\right.\\
 & \left.-\left(\left.v_{x}\right|_{0,\delta y/2}-\left.v_{x}\right|_{-\delta x,\delta y/2}\right)\left(\left.\lambda\right|_{-\delta x/2,\delta y/2}+2\left.\mu\right|_{-\delta x/2,\delta y/2}\right)\right]/\delta x^{2},
\end{aligned}
\label{eq:vx,xx}
\end{equation}
\begin{equation}
\begin{aligned}\left.\left(v_{y,y}\lambda\right)_{,x}\right|_{0,\delta y/2} & =\left[\left(\left.v_{y}\right|_{\delta x/2,\delta y}-\left.v_{y}\right|_{\delta x/2,0}\right)\left.\lambda\right|_{\delta x/2,\delta y/2}\right.\\
 & \quad\left.-\left(\left.v_{y}\right|_{-\delta x/2,\delta y}-\left.v_{y}\right|_{-\delta x/2,0}\right)\left.\lambda\right|_{-\delta x/2,\delta y/2}\right]/\left(\delta x\delta y\right)~,
\end{aligned}
\label{eq:vy,yx}
\end{equation}
and 
\begin{eqnarray*}
\left.\left(\left(v_{x,y}+v_{y,x}\right)\mu\right)_{,y}\right|_{0,\delta y/2} & = & \left(\left(\left.v_{x}\right|_{0,3\delta y/2}-\left.v_{x}\right|_{0,\delta y/2}\right)\left.\mu\right|_{0,\delta y}\right.\\
 &  & \left.-\left(\left.v_{x}\right|_{0,\delta y/2}-\left.v_{x}\right|_{0,-\delta y/2}\right)\left.\mu\right|_{0,0}\right)/\delta y^{2}\\
 &  & +\left(\left(\left.v_{y}\right|_{\delta x/2,\delta y}-\left.v_{y}\right|_{-\delta x/2,\delta y}\right)\left.\mu\right|_{0,\delta y}\right.\\
 &  & \left.-\left(\left.v_{y}\right|_{\delta x/2,0}-\left.v_{y}\right|_{-\delta x/2,0}\right)\left.\mu\right|_{0,0}\right)/\left(\delta x\delta y\right)~.
\end{eqnarray*}
We then define the expression $\partial r_{i}/\partial\left.v_{i}\right|_{x,y}$
as the derivative of the finite difference expression of $r_{i}$
with respect to $\left.v_{i}\right|_{x,y}$. For example, the derivative
of $\big(v_{x,x}\big(\lambda+2\mu\big)\big)_{,x}\bigg|_{0,\delta y/2}$
is

\begin{align*}
\frac{\partial}{\partial\left.v_{x}\right|_{0,\delta y/2}}\left(\big(v_{x,x}\big(\lambda+2\mu\big)\big)_{,x}\bigg|_{0,\delta y/2}\right)= & \left(\left(\left.\lambda\right|_{\delta x/2,\delta y/2}+2\left.\mu\right|_{\delta x/2,\delta y/2}\right)\right.\\
 & \left.-\left(\left.\lambda\right|_{-\delta x/2,\delta y/2}+2\left.\mu\right|_{-\delta x/2,\delta y/2}\right)\right)/\delta x^{2}\,.
\end{align*}

\begin{figure}
\begin{centering}
\includegraphics[width=0.5\columnwidth]{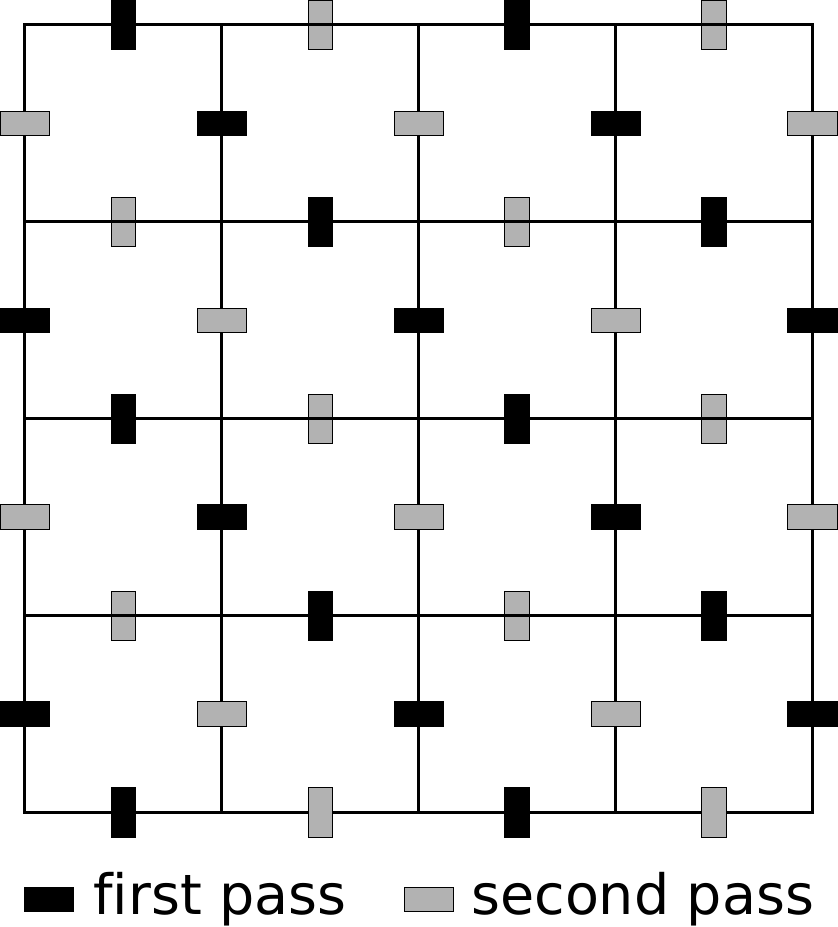}
\par\end{centering}
\caption{Update schedule for Gauss-Seidel relaxation in 2D. Updates for 3D
follow a similar pattern.\label{fig:Red-Black-passes}}

\end{figure}

The Gauss-Seidel update is then given by 

\begin{equation}
\left(\left.v_{i}\right|_{x,y}\right)_{\text{new}}=\left.v_{i}\right|_{x,y}-\frac{r_{i}}{\partial r_{i}/\partial\left.v_{i}\right|_{x,y}}~.\label{eq:v_update}
\end{equation}
 We perform the update in-place in two separate passes as seen in
Figure \ref{fig:Red-Black-passes}. Our discretisation allows us to
update each point within a pass independently of each other. Parallelizing
the method involves partitioning the mesh into regions that each belong
to a different processor. Synchronization only happens before each
pass, where each region gets updates to a single layer of ghost zones.

\subsection{Treatment of Internal Dislocations\label{subsec:Fault-Corrections}}

\subsubsection{Theory\label{subsec:Theory}}

We define faults as a finite-sized internal surfaces where there is
a displacement discontinuity called slip. Fault slip is often described
in piece-wise fault segments where displacement is uniform \cite{okada85,okada92,wang+03a,meade07a,barbot&fialko10a,gimbutas+12,nikkhoo2015triangular},
and we follow this convention. This means that a model of a realistic
fault will be made up of hundreds of fault segments, each with their
own slip. Internal dislocations can cause stress and displacement
singularities at the edges of these segments \cite{paris+65,tada+00,burgmann+94}.
These singularities do not manifest themselves in real earthquakes
because the rock behaves nonlinearly beyond a certain stress by, for
example, breaking. However, the nonlinear behavior occurs over a length
scale that is orders of magnitude smaller than the rest of the model.
So the stress can still get quite high, and these stress concentrations
are key to understanding localized deformation. So modeling algorithms
must not break down in the presence of these singularities.

\begin{figure}
\begin{centering}
\includegraphics[width=0.3\paperwidth]{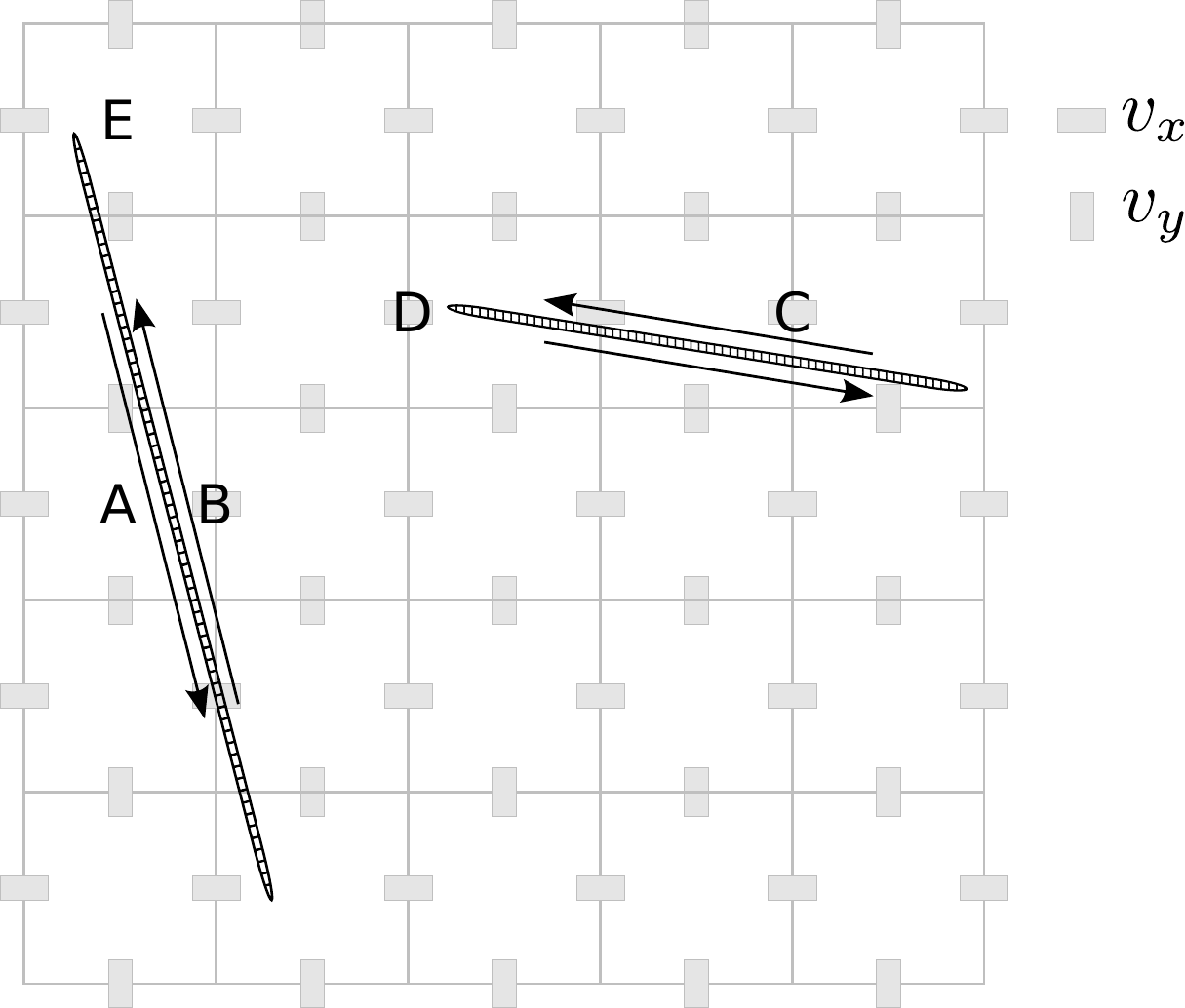} 
\par\end{centering}
\protect\caption{\label{fig:Fault-corrections}Fault corrections on a grid. The stencil
for the derivative $v_{x,x}$ crosses the fault at point A but misses
the fault at point E. The stencil for the derivative $v_{y,yx}$ at
point C crosses the fault but at point D only partially crosses it. }
\end{figure}

To illustrate the method, consider the single faults in 2D in Figure
\ref{fig:Fault-corrections}. The slip $\overrightarrow{s}=\left(s_{x},s_{y}\right)$
on the faults is given as an input to the problem. To compute $v_{x,x}$
at point $A=\left(A_{x},A_{y}\right)$, we would ordinarily write
the finite difference expression 
\[
\left.v_{x,x}\right|_{\text{FD}}=\left(\left.v_{x}\right|_{A_{x}+\delta x/2,A_{y}}-\left.v_{x}\right|_{A_{x}-\delta x/2,A_{y}}\right)/\delta x.
\]
If $v_{x}$ is constant on each side $\left(v_{\text{right}},v_{\text{left}}\right)$,
then the slip $s_{x}$ is the difference between them $s_{x}=v_{\text{right}}-v_{\text{left}}$.
The finite difference then becomes 
\[
\left.v_{x,x}\right|_{\text{FD}}=\left(v_{\text{right}}-v_{\text{left}}\right)/\delta x=s_{x}/\delta x.
\]

This goes to infinity as the resolution improves and $h$ decreases.
However, the true value of $v_{x,x}$ at that point is zero because
$v_{x}$ is constant. The core idea of the original IIM paper \cite{leveque&li94}
is to model these discontinuities explicitly. Then we compute corrections
to apply when computing derivatives. In this case, we can compute
the correct derivative by carefully subtracting away the divergent
term $s_{x}/\delta x$. Then the corrected expression is 
\[
\left.v_{x,x}\right|_{\text{corrected}}=\left.v_{x,x}\right|_{\text{FD}}-s_{x}/\delta x=\left[\left.v_{x}\right|_{A_{x}+\delta x/2}-\left.v_{x}\right|_{A_{x}-\delta x/2}\right]/\delta x-s_{x}/\delta x.
\]
One important note is that this correction is only applied if the
line between $\left.v\right|_{x+\delta x/2}$ and $\left.v\right|_{x-\delta x/2}$
crosses the fault. If it barely misses the fault as in the case at
point E in Figure \ref{fig:Fault-corrections}, there is no correction.
This is a significant difference from other methods such as extended
finite elements, which can have difficulties arising from small cell
volumes or bad aspect ratios \cite{coon+11}. This also implies that
the tip of the fault, as seen by these corrections, is only determined
up to $O(h)$.

When looking at terms with second derivatives, we build them out of
first derivatives. Since the slip is constant along the fault element,
there is no correction in the derivatives, only in the displacements.
This means that we can build $\Delta\left(v_{x,xx}\right)$, the correction
for $v_{x,xx}$, out of $\Delta\left(v_{x,x}\right)$, the corrections
for $v_{x,x}$. In the reference cell, this is 
\begin{equation}
\left.\Delta\left(v_{x,xx}\right)\right|_{0,\delta y}=\left[\left.\Delta\left(v_{x,x}\right)\right|_{\delta x/2,\delta y}-\left.\Delta\left(v_{x,x}\right)\right|_{-\delta x/2,\delta y}\right]/\delta x\label{eq:delta_vxxx}
\end{equation}

To be concrete, when applying this method to Eq. \ref{eq:vx,xx},
the correction at point B in Figure \ref{fig:Fault-corrections} is
\begin{equation}
\left.\Delta\left(\left(\left(\lambda+2\mu\right)v_{x,x}\right)_{,x}\right)\right|_{B}=-\left[\left.\lambda\right|_{B_{x}-\delta x/2,B_{y}}+2\left.\mu\right|_{B_{x}-\delta x/2,B_{y}}\right]\frac{s_{x}}{\delta x^{2}}~.\label{eq:Correction-B}
\end{equation}
The correction to Eq.~\ref{eq:vy,yx} at point C is 
\[
\left.\Delta\left(\left(v_{y,y}\lambda\right)_{,x}\right)\right|_{C}=-\left[\left.\lambda\right|_{C_{x}+\delta x/2,C_{y}}-\left.\lambda\right|_{C_{x}-\delta x/2,C_{y}}\right]\frac{s_{y}}{\delta x\delta y}~,
\]
which is zero if the modulus $\lambda$ is constant. In contrast,
the correction at point D, near the tip of the fault, is

\[
\left.\Delta\left(\left(v_{y,y}\lambda\right)_{,x}\right)\right|_{D}=-\left(\left.\lambda\right|_{D_{x}+\delta x/2,D_{y}}\right)\frac{s_{y}}{\delta x\delta y}~,
\]
because only the derivative 
\[
\left.v_{y,y}\right|_{D_{x}+\delta x/2,D_{y}}
\]
crosses the fault. Finally, the correction to Eq.~\ref{eq:vy,yx}
at point B is zero because each individual correction $\Delta\left(v_{y,y}\right)$
is zero.

Note that these corrections do not depend on the type of slip on the
fault. For example, if the slip has a tensile opening component, the
corrections would have the same form. The only restriction is that
the two sides of the fault must be in contact. With that said, we
have only tested slip along the faults, so we can only speak with
certainty about that kind of slip, referred to as mode II and III
in fracture mechanics.

Excluding the tips, these corrections are exact for the type of slip
being modeled. This means that the stress is consistent and well behaved
across the fault. We might also expect that it would lead to a scheme
that converges as $O\left(h^{2}\right)$. However, the method's uncertainty
about the location of the tips introduces a global error that converges
as $O\left(h\right)$. At the fault tips themselves, the logarithmic
singularity introduces a local error that does not converge.

The above treatment describes a single fault. Since the problem is
linear, we can handle multiple faults, each made up of multiple fault
segments, by adding all of the corrections from individual fault segments
together. This includes the cases where fault segments intersect.

\subsubsection{Implementation}

\begin{figure}[h]
\begin{centering}
\includegraphics[width=0.65\columnwidth]{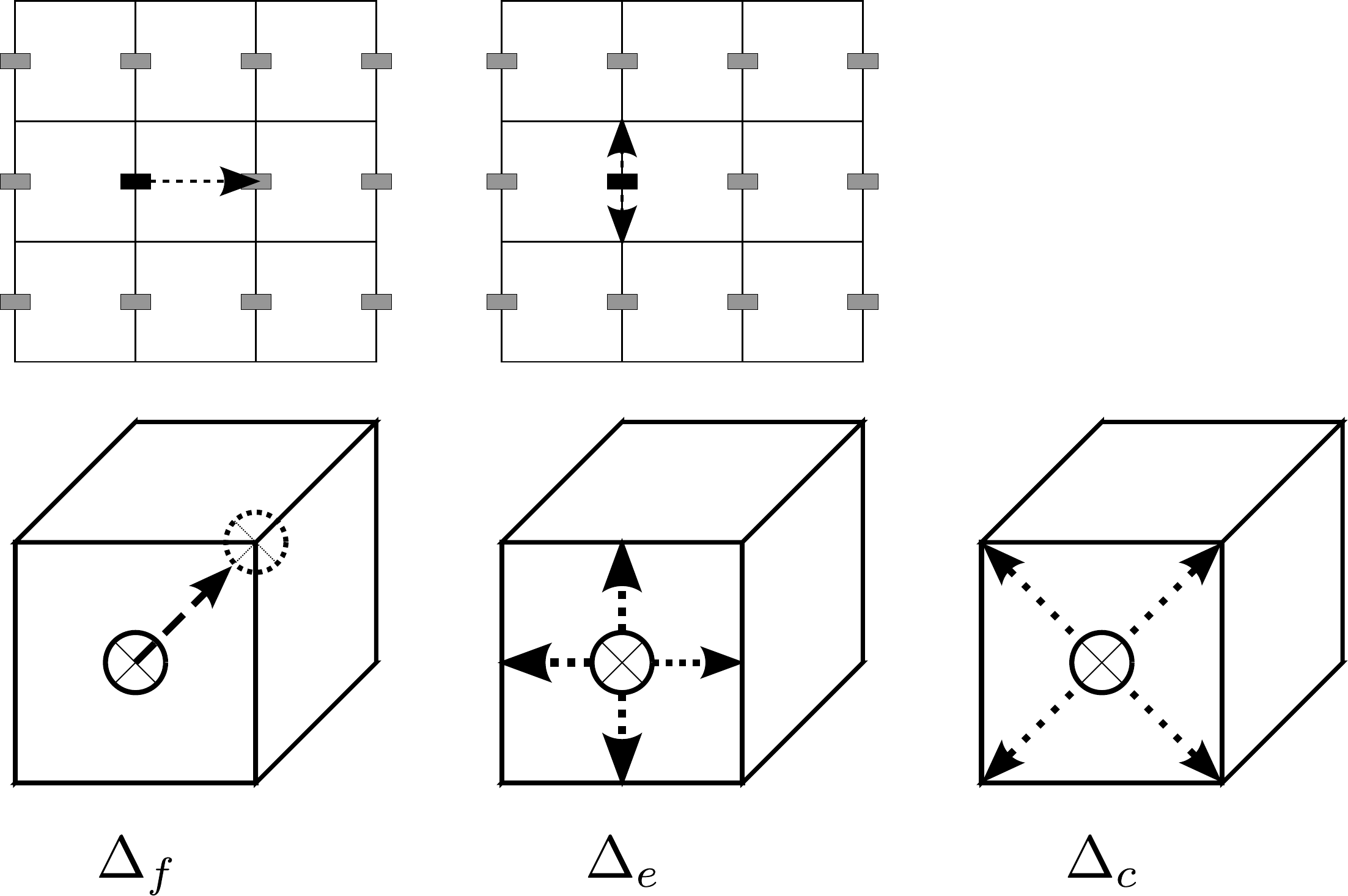} 
\par\end{centering}
\protect\caption{\label{fig:Correction-types}Types of corrections stored. We store
the jump across the cell ($\Delta_{f}$), from the face to the edge
($\Delta_{e}$), and, in 3D, from the face to the corner ($\Delta_{c}$). }
\end{figure}

These corrections do not depend on the computed displacement field.
In that sense, they could be interpreted as body forces $f_{i}$ in
equation \ref{eq:elasto-statics}. In 3D, this would only require
3 additional numbers per cell. However, that analogy breaks down when
we consider the corrections needed when interpolating between coarse
and fine levels for multigrid (Section \ref{subsec:Multigrid}). With
that in mind, we precompute and store the jump in several different
directions as shown in Figure \ref{fig:Correction-types}. In 2D,
we store the jump across a cell ($\Delta_{f}$) and the jump to the
corner ($\Delta_{e}$). Then, for example, the correction in Eq.~\ref{eq:Correction-B}
becomes 
\[
\left.\Delta\left(\left(\left(\lambda+2\mu\right)v_{x,x}\right)_{,x}\right)\right|_{B}=-\left(\left.\lambda\right|_{B_{x}-\delta x/2,B_{y}}+2\left.\mu\right|_{B_{x}-\delta x/2,B_{y}}\right)\frac{\left.\Delta_{fx}\right|_{B_{x}-\delta x/2,B_{y}}}{\delta x^{2}}.
\]

In 2D, this requires storing 
\[
2\left(\Delta_{fx},\Delta_{fy}\right)+4\left(\Delta_{ex+},\Delta_{ex-},\Delta_{ey+},\Delta_{ey-}\right)=6
\]
extra numbers per cell in addition to the 6 ($v_{x}$, $v_{y}$, $\lambda$,
$\mu$, $f_{x}$, $f_{y}$) already needed. In 3D, we store the jump
across the cell ($\Delta_{f}$), from the cell face to the edge ($\Delta_{e}$),
and from the cell face to the corner ($\Delta_{c}$). This requires
\begin{align*}
 & 3\left(\Delta_{fx},\Delta_{fy},\Delta_{fz}\right)\\
 & +12\left(\Delta_{ex+z},\Delta_{ex-z},\Delta_{ey+z},\Delta_{ey-z},\ldots\right)\\
 & +12\left(\Delta_{cx+y+},\Delta_{cx+y-},\Delta_{cx-y+},\Delta_{cx-y-},\ldots\right)=27
\end{align*}
extra numbers per cell in addition to the 9 already needed.

\subsection{Boundary Conditions\label{subsec:Boundary-Conditions}}

We have implemented two different kinds of boundary conditions: Dirichlet,
where the displacement is fixed to a certain value at the boundary,
and stress, where the displacement is set so as to dictate what the
stress is at a point. When imposing these conditions, it turns out
that there is an ordering dependency among the conditions. We must
first impose Dirichlet conditions. Then the shear stress conditions
use values that were just set by the Dirichlet conditions. Finally,
the normal stress conditions use values that were just set by the
Dirichlet and shear stress conditions.

\subsubsection{Dirichlet\label{subsec:Dirichlet_BC}}

\begin{figure}
\begin{centering}
\includegraphics[width=0.3\paperwidth]{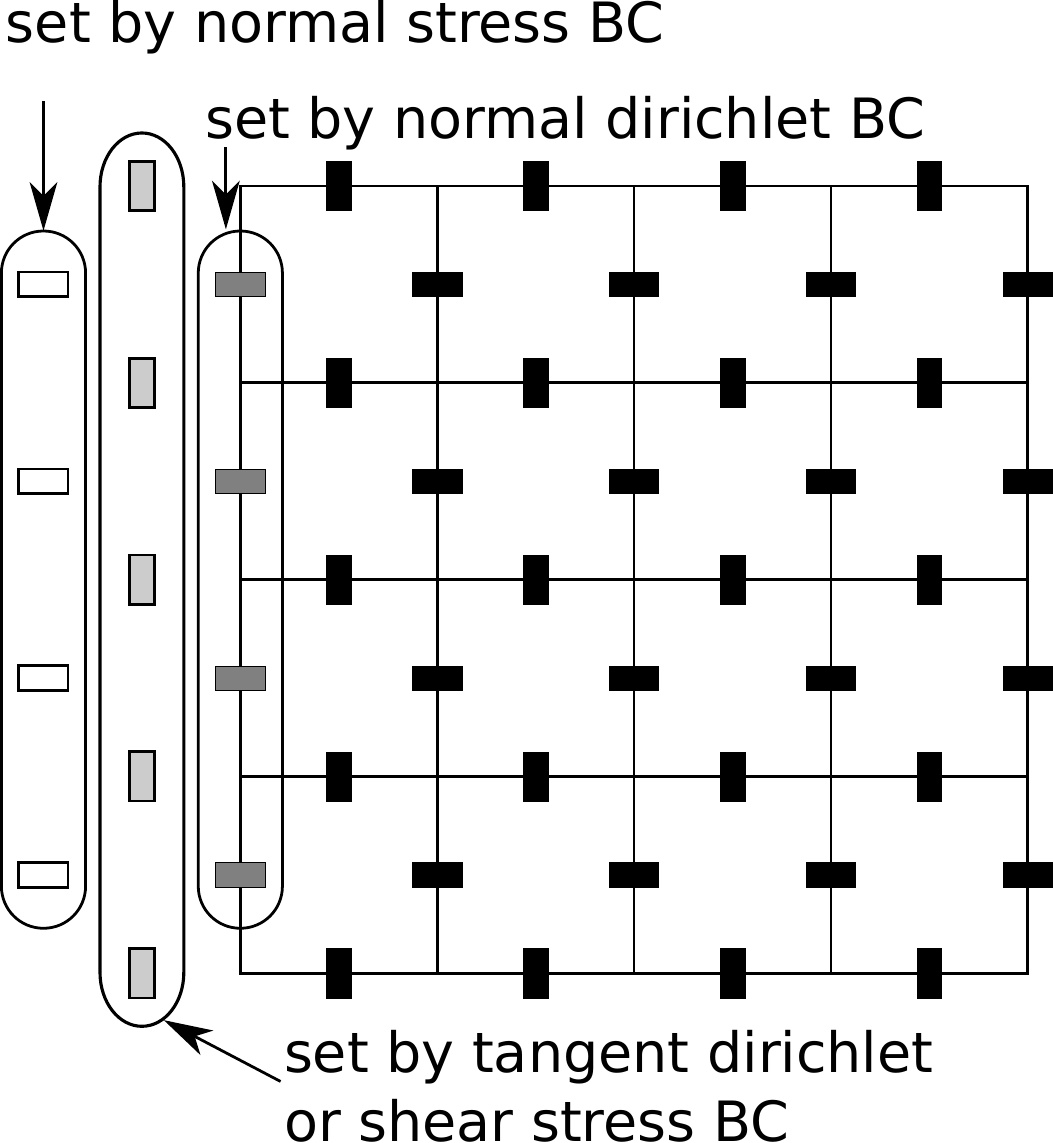} 
\par\end{centering}
\protect\caption{Mapping of points set by the various boundary conditions\label{fig:Boundary-conditions}}
\end{figure}

The simplest boundary condition is Dirichlet conditions on the displacement
normal to the boundary, as shown in Figure \ref{fig:Boundary-conditions}.
In this case, the value at the boundary is simply set to the boundary
value:

\[
v_{x}=v_{\text{BC}}.
\]

For Dirichlet conditions on the displacement tangential to the boundary,
as shown in Figure \ref{fig:Boundary-conditions}, the point outside
is set so that the average of the inner and outer points equal to
the boundary value:

\[
\left.v_{y}\right|_{x-\delta x/2,y}=2v_{\text{BC}}-\left.v_{y}\right|_{x+\delta x/2,y}-\left.\Delta_{ey-}\right|_{x+\delta x/2,y}.
\]
The correction $\left.\Delta_{ey-}\right|_{x+\delta x/2,y}$ is necessary
to handle any faults between $x+\delta x/2$ and $x$. For simplicity,
we define the faults to never extend out of the mesh.

\subsubsection{Stress\label{subsec:Stress_BC}}

A more complicated boundary condition is to set the stress rather
than directly setting the displacement.

\paragraph{Shear Stress}

The $y$ component of the shear stress at an $x$ boundary is

\[
\sigma_{\text{BC}}=\sigma_{xy}=\mu\left(v_{x,y}+v_{y,x}\right).
\]
We apply this condition by setting $v_{y}$ at an outside point

\begin{eqnarray*}
\left.v_{y}\right|_{x-\delta x/2,y} & = & \left.v_{y}\right|_{x+\delta x/2,y}+\left(\left.v_{x}\right|_{x,y+\delta y/2}-\left.v_{x}\right|_{x,y-\delta y/2}\right)\frac{\delta x}{\delta y}-\sigma_{bc}\delta x/\left.\mu\right|_{x,y}\\
 &  & +\left.\Delta_{ey-}\right|_{x+\delta x/2,y}+\left(\left.\Delta_{ex-}\right|_{x,y+\delta y/2}-\left.\Delta_{ex+}\right|_{x,y-\delta y/2}\right)\frac{\delta x}{\delta y}.
\end{eqnarray*}
This depends on $\left.v_{x}\right|_{x,y+\delta y/2}$ and $\left.v_{x}\right|_{x,y-\delta y/2}$,
so the normal Dirichlet condition must be applied before this condition.

\paragraph{Normal Stress}

For the normal stress in the $x$ direction in 2D, the analytic condition
is $\sigma_{\text{BC}}=\sigma_{xx}=2\mu\,v_{x,x}+\lambda\,v_{i,i}$,
which implies 
\[
v_{x,x}=-\frac{\lambda\,v_{y,y}-\sigma_{\text{bc}}}{2\mu+\lambda}~.
\]
We discretize this condition as 
\[
\begin{aligned}\left.v_{x}\right|_{x-\delta x,y+\delta y/2} & =\left.v_{x}\right|_{x+\delta x,y+\delta y/2}-\left.\Delta_{fx}\right|_{x,y+\delta y/2}\\
 & \begin{aligned}+\left[\frac{\lambda_{\text{BC}}}{2\delta y}\right. & \left(\left.v_{y}\right|_{x+\delta x/2,y}+\left.v_{y}\right|_{x-\delta x/2,y}\right.\\
- & \left.\left.\left.v_{y}\right|_{x+\delta x/2,y-\delta y}-\left.v_{y}\right|_{x-\delta x/2,y-\delta y}\right)-\sigma_{\text{BC}}\right]\frac{2\delta x}{\lambda_{\text{BC}}+2\mu_{\text{BC}}}
\end{aligned}
\\
 & -~\left.\Delta_{ey-}\right|_{x+\delta x/2,y-\delta y}\frac{\lambda_{\text{BC}}}{\lambda_{\text{BC}}+2\mu_{\text{BC}}}\frac{\delta x}{\delta y}
\end{aligned}
\]
This interpolates the derivative $v_{y,y}$ onto $\left(x,y+\delta y/2\right)$.
The moduli, $\lambda_{\text{BC}}$ and $\mu_{\text{BC}}$, are also
interpolated there with the usual formula 
\[
\lambda_{\text{BC}}=\frac{1}{2}\big(\left.\lambda\right|_{x,y+\delta y}+\left.\lambda\right|_{x,y}\big)~.
\]
The condition in 3D has an additional term, $v_{z,z}$, which is computed
in a similar manner. This discretization depends on $\left.v_{y}\right|_{x+\delta x/2,y}$,
so the shear stress condition must be applied before this condition.

\subsection{Multigrid on an Adapted Mesh\label{subsec:Multigrid}}

With a smoother (Section \ref{subsec:Gauss-Seidel-Relaxation}), corrections
for faults (Section \ref{subsec:Fault-Corrections}), and boundary
conditons (Section \ref{subsec:Boundary-Conditions}), we can compute
highly accurate solutions to Eq.~\ref{eq:elasto-statics} on a single
grid. This will, however, be very slow. To shorten the time to solution,
we implement adaptive multigrid (Appendix \ref{sec:Adaptive-Multigrid}).
This is essentially an enhancement of the multigrid method for adapted
grids. To implement this, we must first implement coarsening, refinement,
and coarse-fine boundary operators.

\subsubsection{Coarsening\label{subsec:Coarsening}}

\begin{figure}
\begin{centering}
\includegraphics[width=0.2\paperwidth]{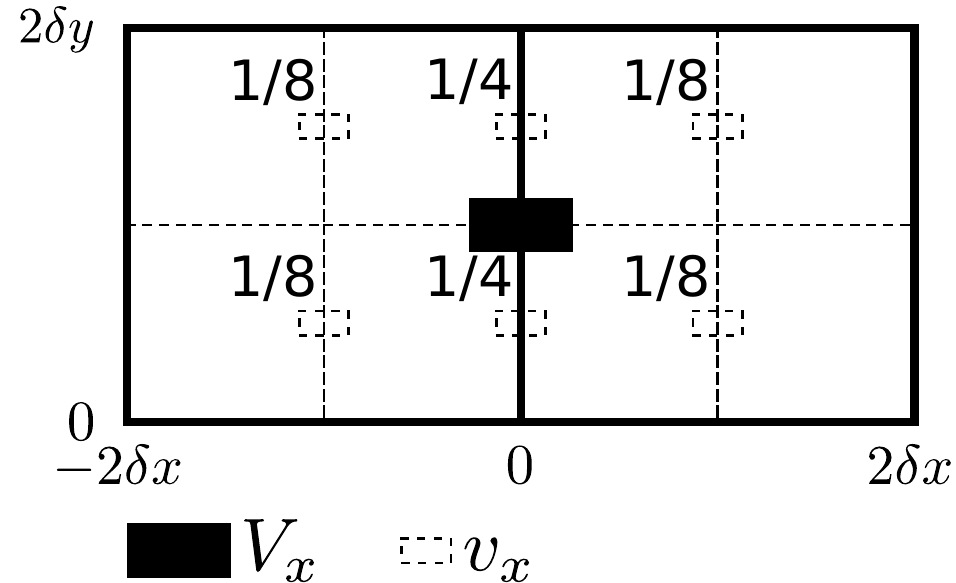} 
\par\end{centering}
\protect\caption{Stencil and weights used for coarsening in 2D\label{fig:Coarsening}}
\end{figure}

Following Albers \cite{albers00} we use weighted arithmetic averages
to coarsen the face centered displacement and residuals. Figure \ref{fig:Coarsening}
shows the fine values used to compute the coarse value for $v_{x}$.
The corresponding expression in the reference cell is

\begin{eqnarray*}
\left.V_{x}\right|_{0,\delta y} & = & \left(\left.v_{x}\right|_{-\delta x,\delta y/2}+2\left.v_{x}\right|_{0,\delta y/2}+\left.v_{x}\right|_{\delta x,\delta y/2}\right.\\
 &  & \left.+\left.v_{x}\right|_{-\delta x,3\delta y/2}+2\left.v_{x}\right|_{0,3\delta y/2}+\left.v_{x}\right|_{\delta x,3\delta y/2}\right)/8\\
 &  & +\left(\left.\Delta_{fx}\right|_{-\delta x,\delta y/2}-\left.\Delta_{fx}\right|_{0,\delta y/2}+\left.\Delta_{fx}\right|_{-\delta x,3\delta y/2}-\left.\Delta_{fx}\right|_{0,3\delta y/2}\right)/8\\
 &  & +\left(\left.\Delta_{ex+}\right|_{0,\delta y/2}+\left.\Delta_{ex-}\right|_{0,3\delta y/2}\right)/2.
\end{eqnarray*}
The expression in 3D is a straightforward extension

\[
\begin{aligned}\left.V_{x}\right|_{0,\delta y,\delta z}= & \frac{1}{16}\left[\left.v_{x}\right|_{-\delta x,\delta y/2,\delta z/2}+2\left.v_{x}\right|_{0,\delta y/2,\delta z/2}+\left.v_{x}\right|_{\delta x,\delta y/2,\delta z/2}\right.\\
 & \qquad+\left.v_{x}\right|_{-\delta x,3\delta y/2,\delta z/2}+2\left.v_{x}\right|_{0,3\delta y/2,\delta z/2}+\left.v_{x}\right|_{\delta x,3\delta y/2,\delta z/2}\\
 & \qquad+\left.v_{x}\right|_{-\delta x,\delta y/2,3\delta z/2}+2\left.v_{x}\right|_{0,\delta y/2,3\delta z/2}+\left.v_{x}\right|_{\delta x,\delta y/2,3\delta z/2}\\
 & \qquad+\left.\left.v_{x}\right|_{-\delta x,3\delta y/2,3\delta z/2}+2\left.v_{x}\right|_{0,3\delta y/2,3\delta z/2}+\left.v_{x}\right|_{\delta x,3\delta y/2,3\delta z/2}\right]\\
+ & \frac{1}{16}\bigg[\left.\Delta_{fx}\right|_{-\delta x,\delta y/2,\delta z/2}-\left.\Delta_{fx}\right|_{0,\delta y/2,\delta z/2}\\
 & \qquad+\left.\Delta_{fx}\right|_{-\delta x,3\delta y/2,\delta z/2}-\left.\Delta_{fx}\right|_{0,3\delta y/2,\delta z/2}\\
 & \qquad+\left.\Delta_{fx}\right|_{-\delta x,\delta y/2,3\delta z/2}-\left.\Delta_{fx}\right|_{0,\delta y/2,3\delta z/2}\\
 & \qquad+\left.\Delta_{fx}\right|_{-\delta x,3\delta y/2,3\delta z/2}-\left.\Delta_{fx}\right|_{0,3\delta y/2,3\delta z/2}\bigg]\\
+ & \frac{1}{4}\bigg[\left.\Delta_{cx++}\right|_{0,\delta y/2,\delta z/2}+\left.\Delta_{cx-+}\right|_{0,3\delta y/2,\delta z/2}\\
 & \qquad+\left.\Delta_{cx+-}\right|_{0,\delta y/2,3\delta z/2}+\left.\Delta_{cx--}\right|_{0,3\delta y/2,3\delta z/2}\bigg]~.
\end{aligned}
\]
At physical boundaries where not all of the values are available,
we average only over the face. In 2D, the expression is 
\[
\left.V_{x}\right|_{0,\delta y}=\frac{1}{2}\left[\left.v_{x}\right|_{0,\delta y/2}+\left.v_{x}\right|_{0,3\delta y/2}+\left.\Delta_{ex+}\right|_{0,\delta y/2}+\left.\Delta_{ex-}\right|_{0,3\delta y/2}\right],
\]
and in 3D it is 
\[
\begin{aligned}\left.V_{x}\right|_{0,\delta y,\delta z} & =\frac{1}{4}\bigg[\left.v_{x}\right|_{0,\delta y/2,\delta z/2}+\left.v_{x}\right|_{0,3\delta y/2,\delta z/2}\\
 & \qquad\qquad\qquad+\left.v_{x}\right|_{0,\delta y/2,3\delta z/2}+\left.v_{x}\right|_{0,3\delta y/2,3\delta z/2}\bigg]\\
 & +\frac{1}{4}\bigg[\left.\Delta_{cx++}\right|_{0,\delta y/2,\delta z/2}+\left.\Delta_{cx-+}\right|_{0,3\delta y/2,\delta z/2}\\
 & \qquad\qquad\qquad+\left.\Delta_{cx+-}\right|_{0,\delta y/2,3\delta z/2}+\left.\Delta_{cx--}\right|_{0,3\delta y/2,3\delta z/2}\bigg]~.
\end{aligned}
\]

\subsubsection{Refinement\label{subsec:Refinement}}

\begin{figure}
\begin{centering}
\includegraphics[width=0.2\paperwidth]{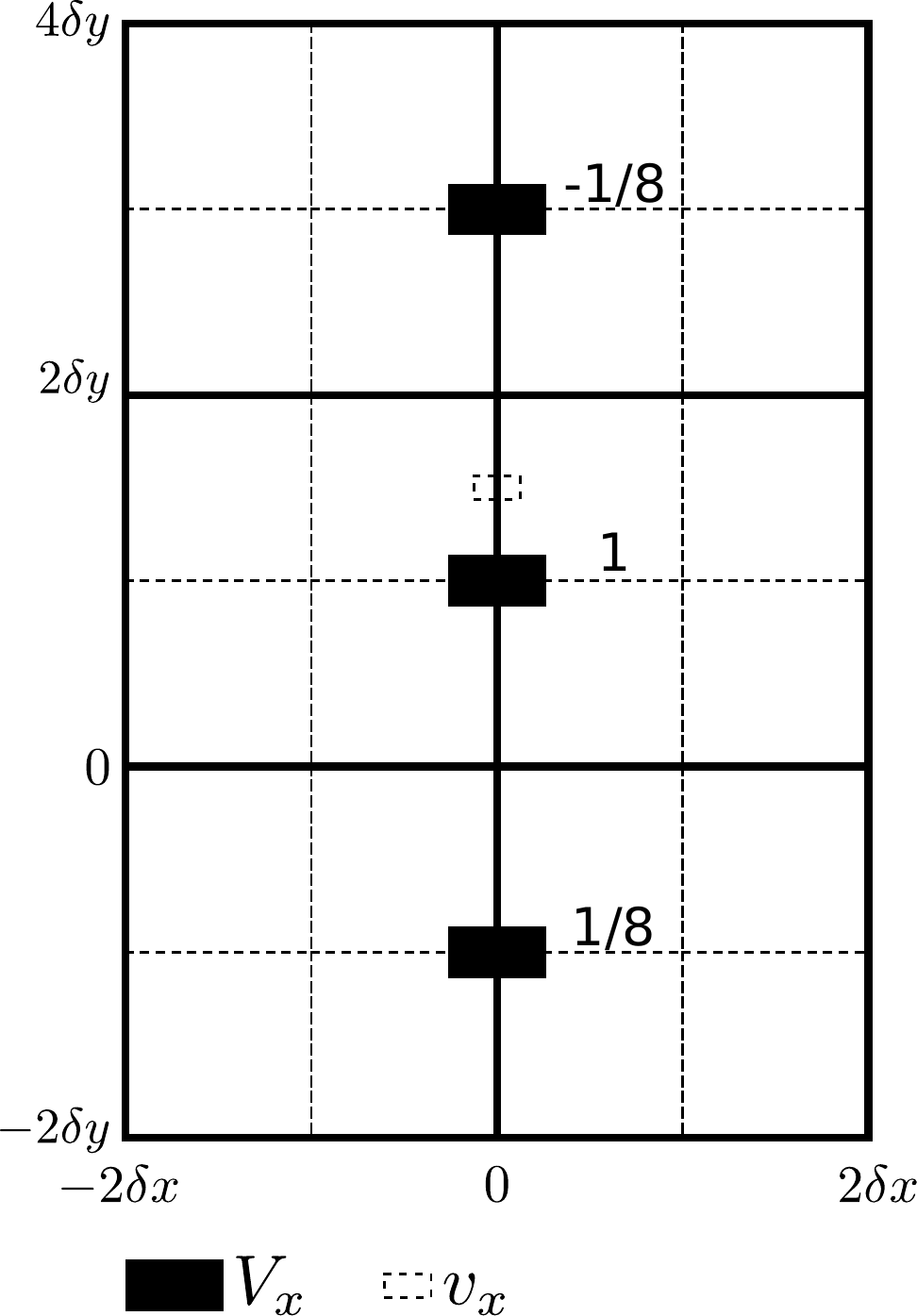} 
\par\end{centering}
\protect\caption{Weights of coarse grid stencil for refining in 2D\label{fig:Refining}}
\end{figure}

To refine the face-centered variables, we use the stencil shown in
Figure \ref{fig:Refining}. We first compute a derivative of the coarse
values, which in 2D is

\[
\left.dV_{x}\right|_{0,\delta y}=\frac{1}{8}\left[\left.V_{x}\right|_{0,3\delta y}-\left.V_{x}\right|_{0,-\delta y}\right]~.
\]
We only refine \textit{corrections} to the displacement, not the displacement
itself. So there is no need to add fault corrections. If we are at
the boundary where one of the variables is not available, we use a
one-sided derivative. For example, at $y=y_{\text{lower}}$, the expression
is 
\[
\left.dV_{x}\right|_{0,y_{\text{lower}}+\delta y}=\frac{1}{4}\left[\left.V_{x}\right|_{0,y_{\text{lower}}+3\delta y}-\left.V_{x}\right|_{0,y_{\text{lower}}+\delta y}\right]~.
\]
The fine value is computed from the closest coarse value and this
computed derivative 
\[
\left.v_{x}\right|_{0,\delta y/2}=\left.V_{x}\right|_{0,\delta y}-\left.dV_{x}\right|_{0,\delta y}~.
\]

In 3D, the expressions look very similar although now we interpolate
along diagonals. For a fine variable on a coarse face, the derivative
is 
\[
\left.dV_{x}\right|_{0,\delta y,\delta z}=\frac{1}{8}\left[\left.V_{x}\right|_{0,3\delta y,3\delta z}-\left.V_{x}\right|_{0,-\delta y,-\delta z}\right]~,
\]
and the fine value is 
\[
\left.v_{x}\right|_{0,\delta y/2,\delta z/2}=\left.V_{x}\right|_{0,\delta y,\delta z}-\left.dV_{x}\right|_{0,\delta y,\delta z}~.
\]
For fine variables in between coarse faces, we average the fine values
on each coarse face: 
\[
\left.v_{x}\right|_{\delta x,\delta y/2,\delta z/2}=\frac{1}{2}\left[\left.v_{x}\right|_{0,\delta y/2,\delta z/2}+\left.v_{x}\right|_{2\delta x,\delta y/2,\delta z/2}\right]~.
\]

\subsubsection{Coarse-Fine Boundaries\label{subsec:Coarse-Fine-boundaries}}

At the interface between coarse and fine levels, we need to compute
boundary conditions for the fine mesh given the coarse surrounding
mesh. There are two cases of coarse-fine boundaries: vector normal
to the interface (e.g., $v_{x}$ at an x=constant boundary), and vector
tangent to the interface (e.g., $v_{x}$ at a y=constant boundary).
When computing these internal boundary conditions, we must use at
least quadratic interpolation to keep the overall error second order
\cite{martin&cartwright96}.

\paragraph{Vector Normal to the Interface}

\begin{figure}
\begin{centering}
\includegraphics[width=0.2\paperwidth]{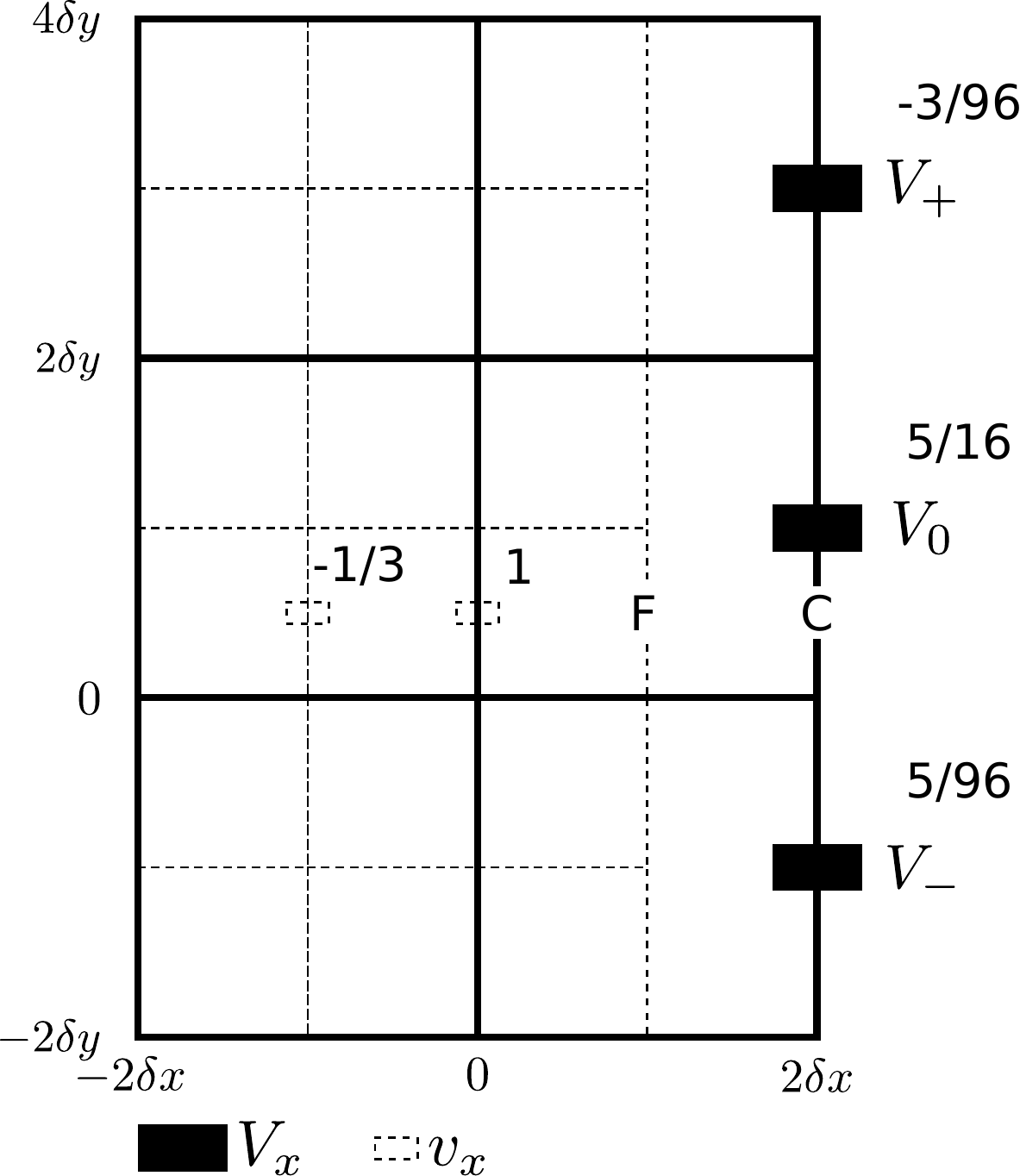} 
\par\end{centering}
\protect\caption{Weights for refining $v_{x}$ to the fine point F at an $x=\text{constant}$
coarse-fine boundary in 2D. The coarse points are first interpolated
to C, then the other fine points are used to quadratically interpolate
to F.\label{fig:coarse_to_fine_2D}}
\end{figure}

Figure \ref{fig:coarse_to_fine_2D} shows the stencil that is used
to compute the fine boundary value on the coarse-fine interface for
the component of a vector that is normal to the interface. The first
step is to interpolate the coarse values to the point C. First, we
define some variables

\begin{eqnarray}
V_{+} & = & \left.V_{x}\right|_{2\delta x,3\delta y}+\left.\bar{\Delta}_{ex-}\right|_{2\delta x,3\delta y}-\left.\bar{\Delta}_{ex+}\right|_{2\delta x,\delta y}\nonumber \\
V_{0} & = & \left.V_{x}\right|_{2\delta x,\delta y}\nonumber \\
V_{-} & = & \left.V_{x}\right|_{2\delta x,-\delta y}+\left.\bar{\Delta}_{ex+}\right|_{2\delta x,-\delta y}-\left.\bar{\Delta}_{ex-}\right|_{2\delta x,\delta y}\nonumber \\
\Delta_{V_{0}} & = & -\left.\bar{\Delta}_{fx}\right|_{0,\delta y}-\left.\Delta_{ex+}\right|_{0,\delta y/2}.\label{eq:V_plus_minus_2D}
\end{eqnarray}
where $\bar{\Delta}$ are the corrections on the coarse grid. Then
the coarse value at C is 
\begin{eqnarray}
dV_{+} & = & V_{+}-V_{0}\nonumber \\
dV_{-} & = & V_{0}-V_{-}\nonumber \\
\left.V_{x}\right|_{C} & = & V_{0}-\left(5dV_{-}+3dV_{+}\right)/32.\label{eq:quad_offset_interpolation}
\end{eqnarray}
The final step is to interpolate along a line to get the fine value
at F

\begin{eqnarray}
\left.v_{x}\right|_{\delta x,\delta y/2} & = & \left.v_{x}\right|_{0,\delta y/2}+\left.\Delta_{fx}\right|_{0,\delta y/2}\nonumber \\
 &  & +\frac{1}{3}\left[\left.V_{x}\right|_{C}-\left.v_{x}\right|_{-\delta x,\delta y/2}+\Delta_{V_{0}}-\left.\Delta_{fx}\right|_{-\delta x,\delta y/2}\right]~.\label{eq:v_fine_interpolate}
\end{eqnarray}

\begin{figure}
\begin{centering}
\includegraphics[width=0.2\paperwidth]{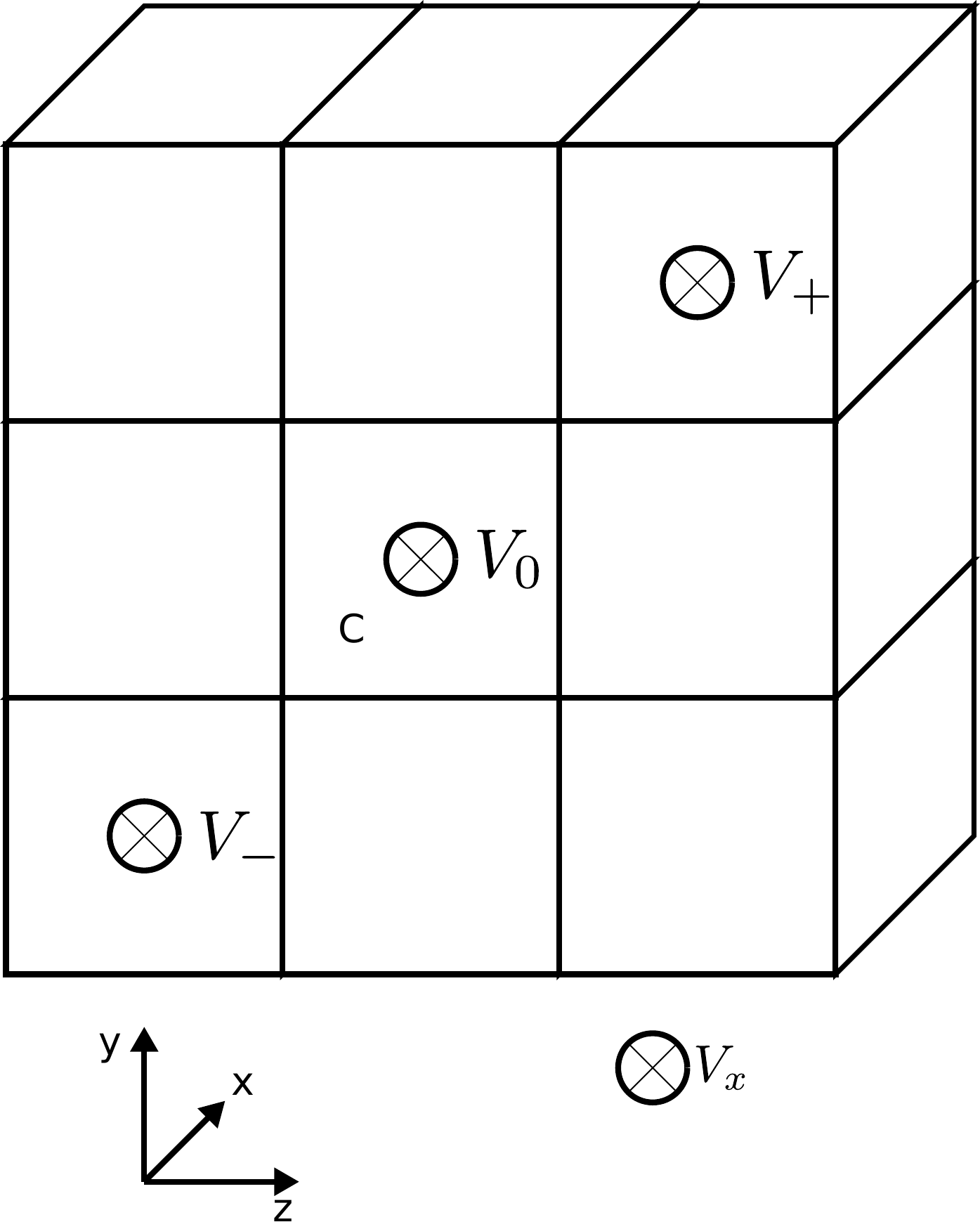} 
\par\end{centering}
\protect\caption{Coarse point part of the stencil for refining $v_{x}$ at the $x=\text{constant}$
coarse-fine boundary in 3D. The $x$ direction is into the picture.
The coarse points are first interpolated to C so as to line up with
the fine points.\label{fig:coarse-fine-3D-cube}}
\end{figure}

In 3D, the interpolation for coarse values is along diagonal directions
as shown in Figure \ref{fig:coarse-fine-3D-cube}. That means that
we can replace Eq.~\ref{eq:V_plus_minus_2D} with

\begin{eqnarray}
V_{+} & = & \left.V_{x}\right|_{2\delta x,3\delta y,3\delta z}+\left.\bar{\Delta}_{cx--}\right|_{2\delta x,3\delta y,3\delta z}-\left.\bar{\Delta}_{cx++}\right|_{2\delta x,\delta y,\delta z}\nonumber \\
V_{0} & = & \left.V_{x}\right|_{2\delta x,\delta y,\delta z}\nonumber \\
V_{-} & = & \left.V_{x}\right|_{2\delta x,-\delta y,-\delta z}+\left.\bar{\Delta}_{cx++}\right|_{2\delta x,-\delta y,-\delta z}-\left.\bar{\Delta}_{cx--}\right|_{2\delta x,\delta y,\delta z}\nonumber \\
\Delta_{V_{0}} & = & -\left.\bar{\Delta}_{fx}\right|_{0,\delta y}-\left.\Delta_{cx++}\right|_{0,\delta y/2},\label{eq:V_plus_minus_3D}
\end{eqnarray}
and then use Eq.~\ref{eq:quad_offset_interpolation} as is. Eq.~\ref{eq:v_fine_interpolate}
is only slightly modified for 3D 
\begin{equation}
\begin{aligned}\left.v_{x}\right|_{\delta x,\delta y/2,\delta z/2} & =\left.v_{x}\right|_{0,\delta y/2,\delta z/2}+\left.\Delta_{fx}\right|_{0,\delta y/2,\delta z/2}\\
 & \qquad+\frac{1}{3}\left[\left.V_{x}\right|_{C}-\left.v_{x}\right|_{-\delta x,\delta y/2,\delta z/2}+\Delta_{V_{0}}-\left.\Delta_{fx}\right|_{-\delta x,\delta y/2,\delta z/2}\right]~.
\end{aligned}
\label{eq:coarse-fine-3D}
\end{equation}
If one of the coarse points is outside the physical domain, then we
use a simpler interpolation. If $V_{+}$ is outside, then 
\[
\left.V_{x}\right|_{C}=\frac{1}{4}\left[3V_{0}+V_{-}+\left.\Delta_{cx++}\right|_{2\delta x,-\delta y,-\delta z}-\left.\Delta_{cx--}\right|_{2\delta x,\delta y,\delta z}\right]~,
\]
and if $V_{-}$ is outside then

\[
\left.V_{x}\right|_{C}=\frac{1}{4}\left[5V_{0}-V_{+}-\left.\Delta_{cx--}\right|_{2\delta x,3\delta y,3\delta z}+\left.\Delta_{cx++}\right|_{2\delta x,\delta y,\delta z}\right]~.
\]
Eq.~\ref{eq:coarse-fine-3D} is used unchanged.

\paragraph{Vector Tangent to the Interface}

\begin{figure}
\begin{centering}
\includegraphics[width=0.2\paperwidth]{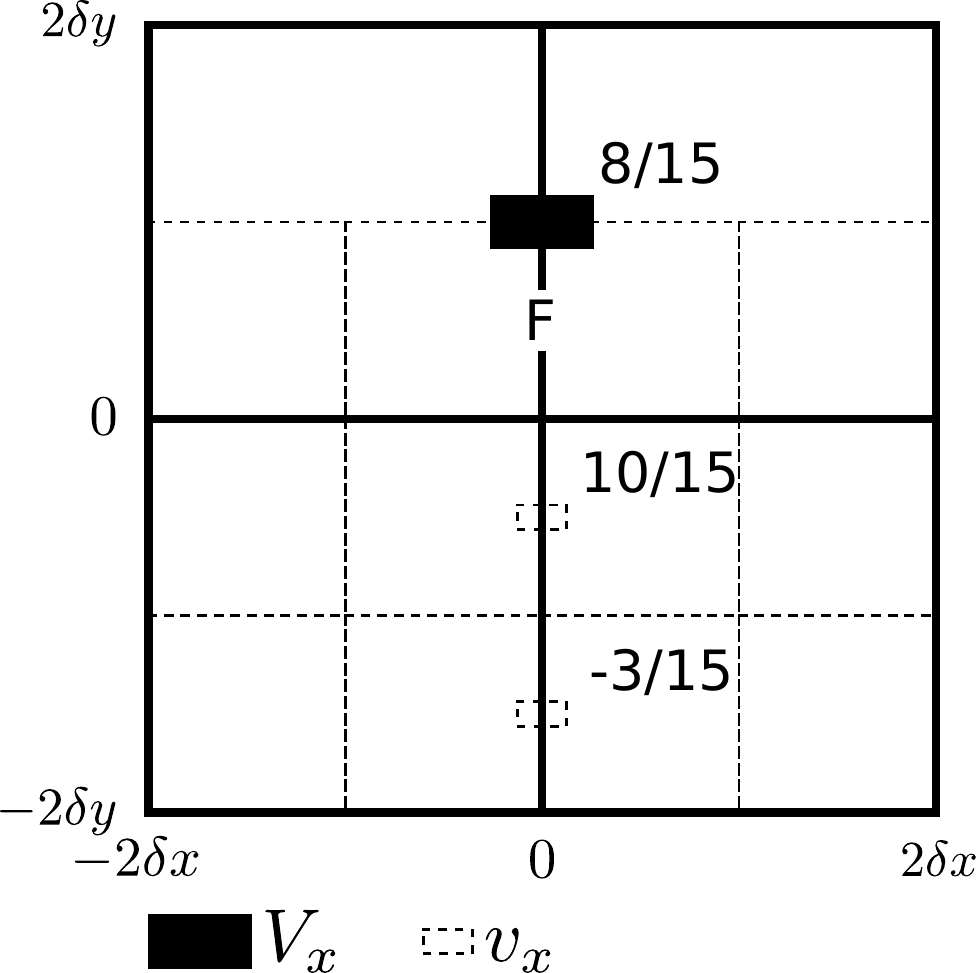} 
\par\end{centering}
\protect\caption{Weights for refining $v_{x}$ to the fine point F at the $y=\text{constant}$
coarse-fine boundary in 2D.\label{fig:coarse-fine-tangent-2D}}
\end{figure}

Figure \ref{fig:coarse-fine-tangent-2D} shows the stencil used for
refinement in 2D when the vector is tangential to the interface. For
the case where the coarse and fine values are on the same coordinate
axis, the interpolation is

\begin{equation}
\begin{aligned}\left.v_{x}\right|_{0,\delta y/2} & =\frac{1}{15}\bigg[8\left.V_{x}\right|_{0,\delta y}+10\left.v_{x}\right|_{0,-\delta y/2}-3\left.v_{x}\right|_{0,-3\delta y/2}\\
 & \quad\qquad+8\left.\bar{\Delta}_{ex-}\right|_{0,\delta y}-8\left.\Delta_{ex+}\right|_{0,\delta y/2}\\
 & \quad\qquad+7\left.\Delta_{ex+}\right|_{0,-\delta y/2}-7\left.\Delta_{ex-}\right|_{0,\delta y/2}\\
 & \quad\qquad-3\left.\Delta_{ex+}\right|_{0,-3\delta y/2}+3\left.\Delta_{ex-}\right|_{0,-\delta y/2}\bigg]~.
\end{aligned}
\label{eq:refine-tangent}
\end{equation}
When the fine value does not lie along the coarse grid, we use a simple
average of the neighboring coarse values 
\[
\left.V_{x}\right|_{\delta x,\delta y}\equiv\frac{1}{2}\left[\left.V_{x}\right|_{0,\delta y}+\left.V_{x}\right|_{2\delta x,\delta y}\right]~,
\]
and the interpolation becomes

\[
\begin{aligned}\left.v_{x}\right|_{\delta x,\delta y/2}=\frac{1}{15}\bigg[ & \left(8\left.V_{x}\right|_{\delta x,\delta y}+10\left.v_{x}\right|_{\delta x,-\delta y/2}-3\left.v_{x}\right|_{\delta x,-3\delta y/2}\right)\\
 & +4\bigg(\left.\bar{\Delta}_{ex-}\right|_{0,\delta y}-\left.\Delta_{ex+}\right|_{0,\delta y/2}+\left.\Delta_{fx}\right|_{0,\delta y/2}\\
 & \qquad+\left.\bar{\Delta}_{ex-}\right|_{2\delta x,\delta y}-\left.\Delta_{ex+}\right|_{2\delta x,\delta y/2}-\left.\Delta_{fx}\right|_{2\delta x,\delta y/2}\bigg)\\
 & +7\left(\left.\Delta_{ex+}\right|_{\delta x,-\delta y/2}-\left.\Delta_{ex-}\right|_{\delta x,\delta y/2}\right)\\
 & -3\left(\left.\Delta_{ex+}\right|_{\delta x,-3\delta y/2}-\left.\Delta_{ex-}\right|_{\delta x,-\delta y/2}\right)\bigg]
\end{aligned}
\]
At the $x=x_{\text{min}}$ or $x=x_{\text{max}}$ corner, some of
the fine corrections (e.g. $\left.\Delta_{fx}\right|_{0,\delta y/2}$)
are not necessarily defined. For the $x=x_{\text{min}}$ boundary,
we work around this by correcting the coarse value at $\left(0,\delta y\right)$
to $\left(2\delta x,\delta y\right)$ first, and then using the same
correction from $\left(2\delta x,\delta y\right)$ to $\left(\delta x,\delta y/2\right)$.
With this, the interpolation becomes

\[
\begin{aligned}\left.v_{x}\right|_{\delta x,\delta y/2}=\frac{1}{15}\bigg[ & \left(8\left.V_{x}\right|_{\delta x,\delta y}+10\left.v_{x}\right|_{\delta x,-\delta y/2}-3\left.v_{x}\right|_{\delta x,-3\delta y/2}\right)\\
 & +4\left(\left.\bar{\Delta}_{fx}\right|_{0,\delta y}+2\left(\left.\bar{\Delta}_{ex-}\right|_{2\delta x,\delta y}-\left.\Delta_{ex+}\right|_{2\delta x,\delta y/2}-\left.\Delta_{fx}\right|_{2\delta x,\delta y/2}\right)\right)\\
 & +7\left(\left.\Delta_{ex+}\right|_{\delta x,-\delta y/2}-\left.\Delta_{ex-}\right|_{\delta x,\delta y/2}\right)\\
 & -3\left(\left.\Delta_{ex+}\right|_{\delta x,-3\delta y/2}-\left.\Delta_{ex-}\right|_{\delta x,-\delta y/2}\right)\bigg].
\end{aligned}
\]

\begin{figure}
\begin{centering}
\includegraphics[width=0.2\paperwidth]{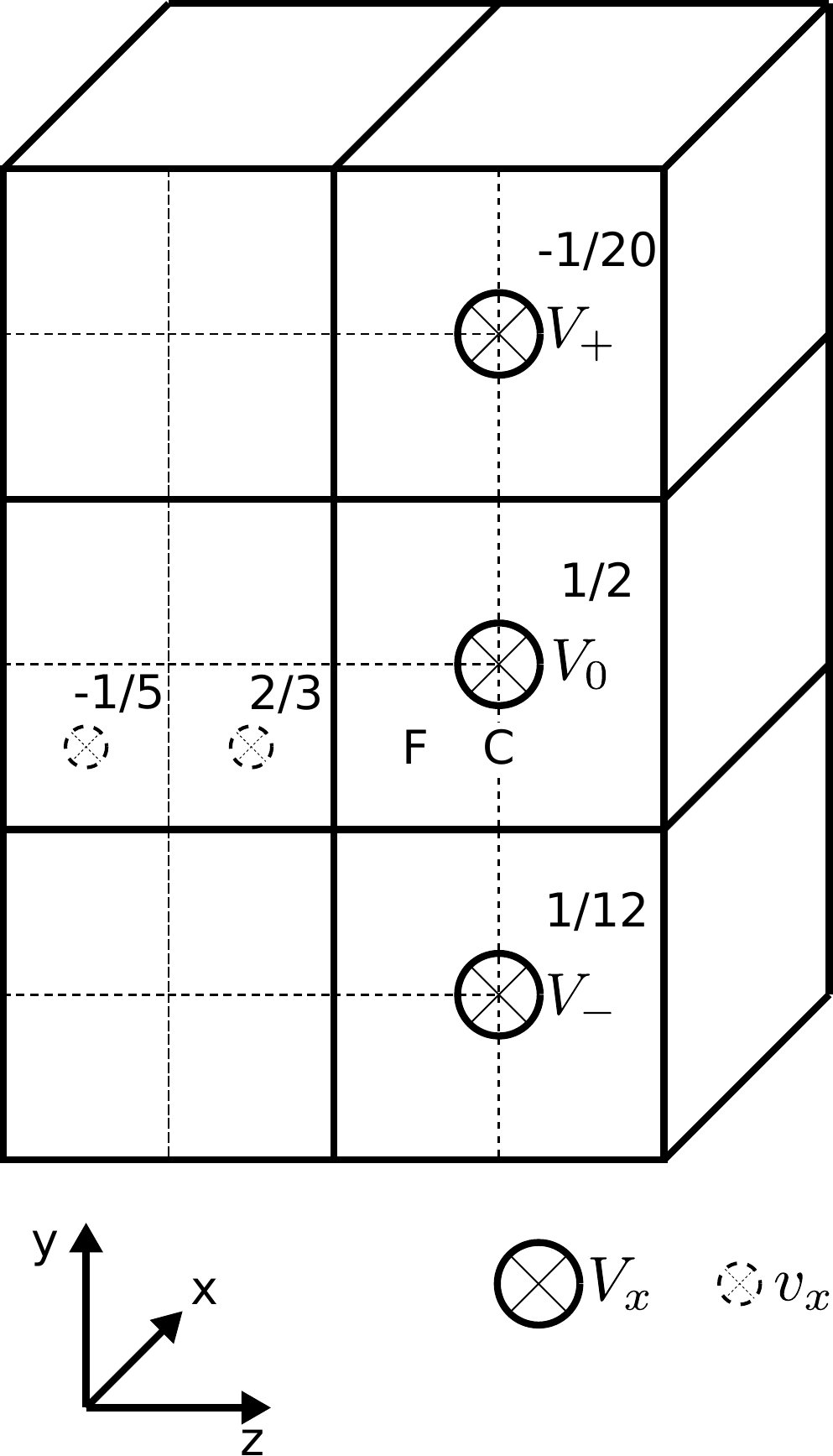} 
\par\end{centering}
\protect\caption{Weights for refining $v_{x}$ to F at the $z=\text{constant}$ coarse-fine
boundary in 3D.\label{fig:coarse-fine-tangent-3D}}
\end{figure}

Figure \ref{fig:coarse-fine-tangent-3D} shows the points used for
refinement in 3D when the coarse and fine values are on the same coordinate
axis. Defining

\begin{eqnarray*}
V_{+} & = & \left.V_{x}\right|_{0,\delta y,3\delta z}+\left.\bar{\Delta}_{ex,z-}\right|_{0,\delta y,3\delta z}-\left.\bar{\Delta}_{ex,z+}\right|_{0,\delta y,\delta z}\\
V_{0} & = & \left.V_{x}\right|_{0,\delta y,\delta z}\\
V_{-} & = & \left.V_{x}\right|_{0,\delta y,-\delta z}+\left.\bar{\Delta}_{ex,z+}\right|_{0,\delta y,-\delta z}-\left.\bar{\Delta}_{ex,z-}\right|_{0,\delta y,\delta z}\\
\Delta_{V_{0}} & = & \left.\bar{\Delta}_{c--}\right|_{0,\delta y,\delta z}-\left.\Delta_{c++}\right|_{0,0,0},
\end{eqnarray*}
we then use Eq.~\ref{eq:quad_offset_interpolation} to compute the
coarse value at point C. Then we use an expression much like \ref{eq:refine-tangent}
to compute the fine value 
\[
\begin{aligned}\left.v_{x}\right|_{0,\delta y/2,\delta z/2}=\frac{1}{15}\bigg[ & 8\left.V_{x}\right|_{C}+10\left.v_{x}\right|_{0,-\delta y/2,\delta z/2}-3\left.v_{x}\right|_{0,-3\delta y/2,\delta z/2}\\
 & +8\Delta_{V_{0}}+7\left.\Delta_{ex,y+}\right|_{0,-\delta y/2,\delta z/2}-3\left.\Delta_{ex,y+}\right|_{0,-3\delta y/2,\delta z/2}\bigg]~.
\end{aligned}
\]

\subsection{Generating the Adapted Mesh\label{subsec:Adaptive-Solution}}

The final part of the method is generating a mesh. Starting with a
uniform grid at the coarsest resolution 
\begin{enumerate}
\item Compute a solution on the current set of grids (section \ref{subsec:Multigrid}). 
\item If the current number of levels is less than the maximum number of
levels

\begin{enumerate}
\item Compute the maximum curvature at each cell center $\left(x+\delta x/2,y+\delta y/2\right)$.
The curvature in the $x$ direction with fault corrections is 
\[
\begin{aligned}\left.C_{x}\right|_{x+\delta x/2,y+\delta y/2} & =\left.v_{x}\right|_{x-\delta x,y+\delta y/2}-\left.v_{x}\right|_{x,y+\delta y/2}\\
 & -\left.v_{x}\right|_{x+\delta x,y+\delta y/2}+\left.v_{x}\right|_{x+2\delta x,y+\delta y/2}\\
 & -\left.\Delta_{fx}\right|_{x-\delta x/2,y+\delta y/2}+\left.\Delta_{fx}\right|_{x+3\delta x/2,y+\delta y/2}.
\end{aligned}
\]
At the boundaries, not all points are defined. For example, at an
$x=x_{\text{lower}}$ Dirichlet boundary, $\left.v_{x}\right|_{x-\delta x,y+\delta y/2}$
may not be defined. In these cases, we use a one-sided curvature 
\[
\begin{aligned}\left.C_{x}\right|_{x_{\text{lower}}+\delta x/2,y+\delta y/2}= & +\left.v_{x}\right|_{x_{\text{lower}},y+\delta y/2}\\
 & -2\left.v_{x}\right|_{x_{\text{lower}}+\delta x,y+\delta y/2}\\
 & +\left.v_{x}\right|_{x_{\text{lower}}+2\delta x,y+\delta y/2}\\
 & -\left.\Delta_{fx}\right|_{x+\delta x/2,y+\delta y/2}\\
 & +\left.\Delta_{fx}\right|_{x+3\delta x/2,y+\delta y/2}.
\end{aligned}
\]
We then compute the maximum curvature 
\[
C_{\text{max}}=\max\left(C_{x},C_{y},C_{z}\right).
\]
\item Refine all cells where $C_{\text{max}}>\epsilon$, where $\epsilon$
is a fixed number, unless the maximum number of mesh refinements has
been reached. Note that $\epsilon$ is an absolute rather than a relative
error. 
\item Recurse back to step 1 with the new set of grids. 
\end{enumerate}
\end{enumerate}
At fault tips, the displacement is singular and so can never be adequately
resolved. However, at a finite distance from the singularity, AMR
solutions can still converge \cite{bai&brandt87}.

\subsection{Accuracy}

When solving equation \ref{eq:elasto-statics} in the presence of
faults, there will always be inaccuracies because of the singularities
at the tips of the faults. Away from the singularity, we expect $O\left(h\right)$
convergence (Section \ref{subsec:Theory}). At the singularity, analysis
becomes difficult because the Taylor series approximation breaks down.
However, the scheme in Section \ref{subsec:Adaptive-Solution} monitors
this error and refines where needed. This means that, where the algorithm
has stopped refining, the discretization error should be less than
the error bound $\epsilon$. In practice, the actual error will be
larger because the local error gets integrated along the points from
the boundaries and singularities.

An additional source of error arises because we only approximately
solve equation \ref{eq:elasto-statics}. If there is an error in the
displacement $\xi_{i}$, that will generate an error in the derivative
$v_{i,j}$ of approximately $\xi_{i}/\delta x$, where $\delta x$
is the grid spacing. This implies that, for a given $\xi_{i}$, the
error in the stress will be at least

\[
\varepsilon\left(\sigma_{ji}\right)\apprge\frac{\xi_{i}\min\left(\lambda,\mu\right)}{\delta x}.
\]
where $\min\left(\lambda,\mu\right)$ is the smallest value of $\lambda$
or $\mu$. The modulus does not, in our problem, vary wildly, so $\nabla\mu\ll\mu/\delta x$.
This implies that the error in the divergence of the stress is approximately
\[
\varepsilon\left(\sigma_{ji,j}\right)\sim\varepsilon\left(\sigma_{ji}\right)/\delta x.
\]
Using equation \ref{eq:residual}, we relate this to the size of the
residual $r_{i}$

\[
r_{i}\sim\varepsilon\left(\sigma_{ji,j}+f_{i}\right).
\]
Errors in $v_{i}$ do not contribute to errors in $f_{i}$, so that
term can be neglected. Simplifying this gives an estimate for the
size of the error $\xi_{i}$ in terms of the residual

\[
\xi_{i}\apprle r_{i}\delta x^{2}/\min\left(\lambda,\mu\right),
\]
This error will become comparable to the discretization error when
$\xi_{i}=\epsilon$, so we can turn this around to find the minimum
resolution required to ensure that the solver error is smaller than
the discretization error

\begin{equation}
\delta x\apprge\sqrt{\min\left(\lambda,\mu\right)\epsilon/r_{i}}.\label{eq:minimum_dx}
\end{equation}

To be clear, this analysis only covers errors in solving \ref{eq:elasto-statics}
using fault segments. We do not claim to model all of the physical
effects (e.g. non-linear rheologies, topography, curved faults).

\section{Analytic Tests}

We have implemented this method in the parallel, adaptive code Gamra.
Gamra uses the SAMRAI framework \cite{hornung+02,hornung+06} to handle
the bookkeeping associated with multiple levels, multiple grids, and
multiple parallel processes. SAMRAI is a mature, freely available,
actively developed framework for large-scale parallel structured adaptive
mesh refinement. SAMRAI uses MPI to coordinate work among the different
processors. This has allowed us to run Gamra on a wide variety of
parallel architectures: SMP nodes, traditional Linux clusters, a Blue
Gene /Q, and the Intel Xeon Phi 5110p GPGPU.

In this section we perform a number of tests to ensure that the algorithm
works as expected and that we implemented it correctly. We have verified
that the code works in both 2D and 3D, but mostly discuss the 3D results
for brevity. The tests are available from the Gamra repository\footnote{\href{https://bitbucket.org/wlandry/gamra}{https://bitbucket.org/wlandry/gamra},
changeset \texttt{679:c8843527b10f18758e58011c57d5aa61098c88e2}, directory
\texttt{input/benchmarks/Elastic}}.

\subsection{Expanding Cylinder in a Heterogeneous Medium}

\begin{figure}
\centering{}\includegraphics[width=0.42\columnwidth]{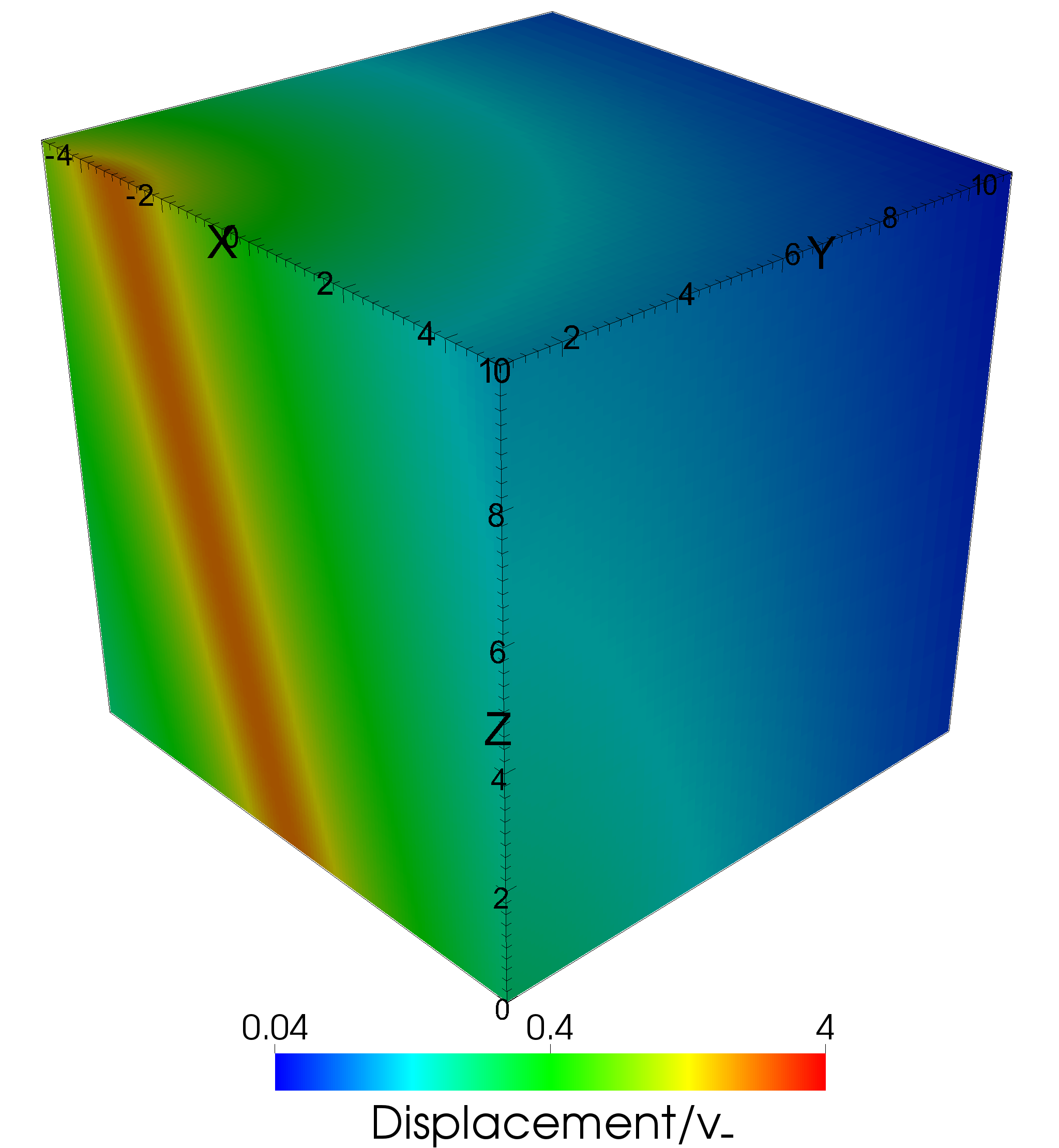}\includegraphics[width=0.45\columnwidth]{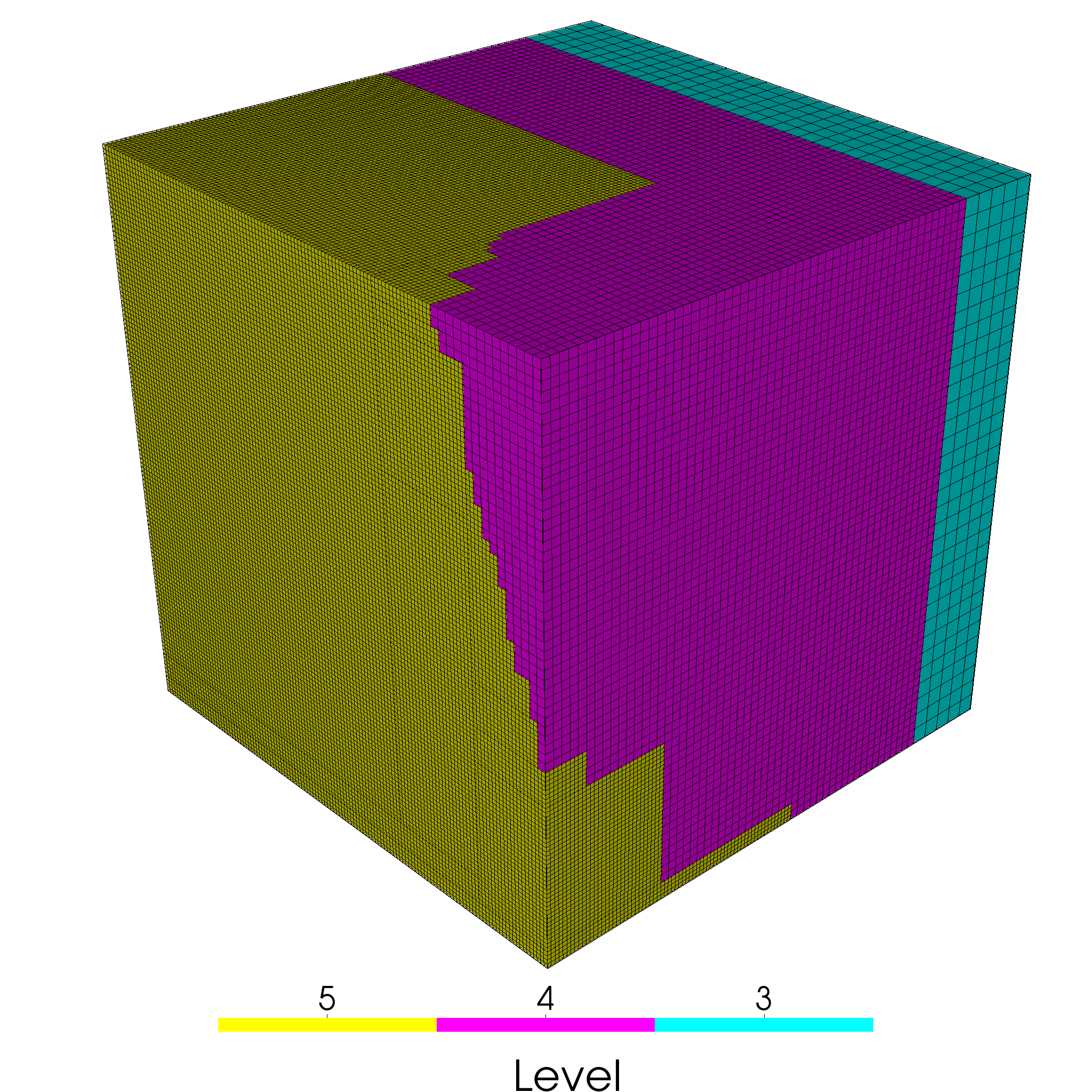}\protect\caption{A cutout of the scaled displacement magnitude of a computed solution
and its associated adapted mesh levels for an expanding cylinder in
3D. The axis of the cylinder is angled 18 degrees from the $x$ axis.
The model covers $\left(-5,1,0\right)$ to $\left(5,11,10\right)$.
The offset is to avoid the singularity at the origin. The boundary
conditions, set from the analytic solution, are Dirichlet for the
normal components \ref{subsec:Dirichlet_BC} and stress for the tangent
components \ref{subsec:Stress_BC}. The equivalent resolution is $128\times128\times128$.
\label{fig:cylinder3D_solution}}
\end{figure}

This is a simple test to ensure that we handle variable elastic modulus
correctly. In cylindrical symmetry, if we set the moduli and body
forces to 
\begin{align*}
\rho= & \sqrt{x^{2}+y^{2}}\\
\mu= & \mu_{0}\rho\\
\lambda= & \frac{2}{3}\mu\\
f_{i}= & 0\,,
\end{align*}
 then the basis functions for solutions to Eq. \ref{eq:elasto-statics}
which are purely cylindrical with no rotation or vertical components
are

\begin{align}
v_{\rho}= & v_{_{-}}\rho^{-3/2}+v_{+}\rho^{1/2}\nonumber \\
v_{\theta}= & 0\nonumber \\
v_{z}= & 0\,.\label{eq:expanding_cylinder}
\end{align}
 To make the test more rigorous, we rotate the solution by an angle
$\theta$ around the $y$ axis. Figure \ref{fig:cylinder3D_solution}
shows a numerical solution and its associated adapted grid for a model
with $\mu_{0}=1.4$, $v_{-}=1$, $v_{+}=0$, and $\theta=18^{\circ}$.
Table \ref{tab:Cylinder_error} shows the $L^{1}$, $L^{2}$, and
$L^{\infty}$ error in $v_{x}$. While the $L^{1}$ and $L^{2}$ errors
do converge, they do not converge as $O\left(h^{2}\right)$. The error
in the unrefined regions no longer decreases, because the mesh does
not get smaller there. The integral of these small errors over the
large unrefined volume is large enough to affect the overall convergence
rate. This is in contrast to the $L^{\infty}$ error, which converges
uniformly at the expected $O\left(h^{2}\right)$ rate.

\begin{table}
\begin{centering}
\begin{tabular}{|c||c|c|c|c|}
\hline 
Level  & $L^{1}$  & $L^{2}$  & $L^{\infty}$  & $L_{n}^{\infty}/L_{n+1}^{\infty}$\tabularnewline
\hline 
\hline 
0  & 58.1 & 5.10 & 0.859 & \tabularnewline
\hline 
1  & 5.86 & 0.612 & 0.168 & 5.11\tabularnewline
\hline 
2  & 1.36 & 0.125 & 0.0409 & 4.11\tabularnewline
\hline 
3  & 0.344 & 0.0313  & 0.0118 & 3.48\tabularnewline
\hline 
4  & 0.0819 & 7.79e-3 & 3.13e-3 & 3.76\tabularnewline
\hline 
5  & 0.0378  & 2.33e-3 & 8.02e-4 & 3.90\tabularnewline
\hline 
\end{tabular}
\par\end{centering}
\protect\caption{\label{tab:Cylinder_error}$L^{1}$, $L^{2}$, and $L^{\infty}$ errors
and $L^{\infty}$ convergence rate in $v_{x}$ at different maximum
refinement levels for the 3D expanding cylinder.}
\end{table}

\subsection{Internal dislocations \label{subsec:Okada-analytic}}

\begin{figure}
\centering{}\includegraphics[width=0.45\columnwidth]{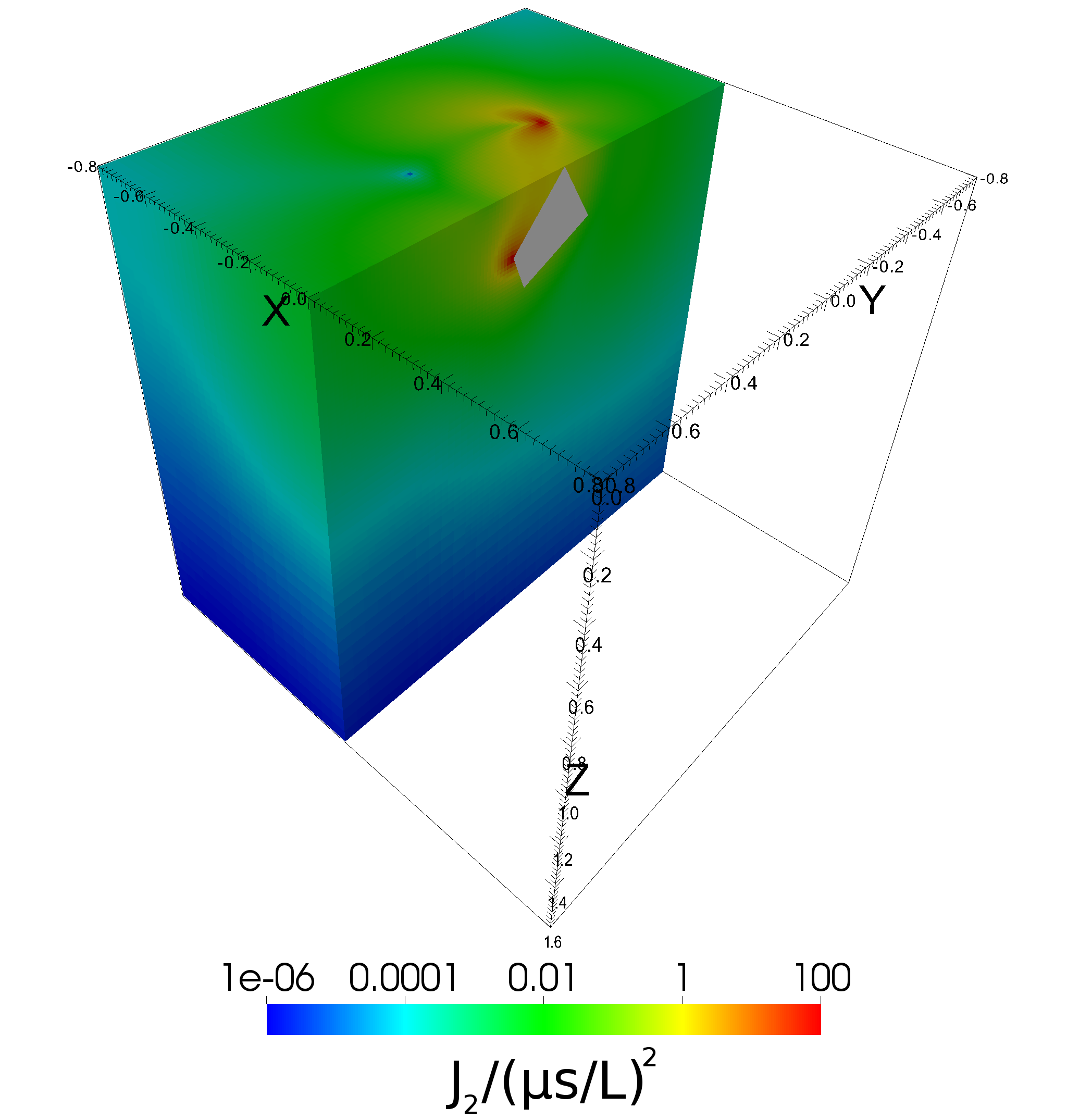}\includegraphics[width=0.45\columnwidth]{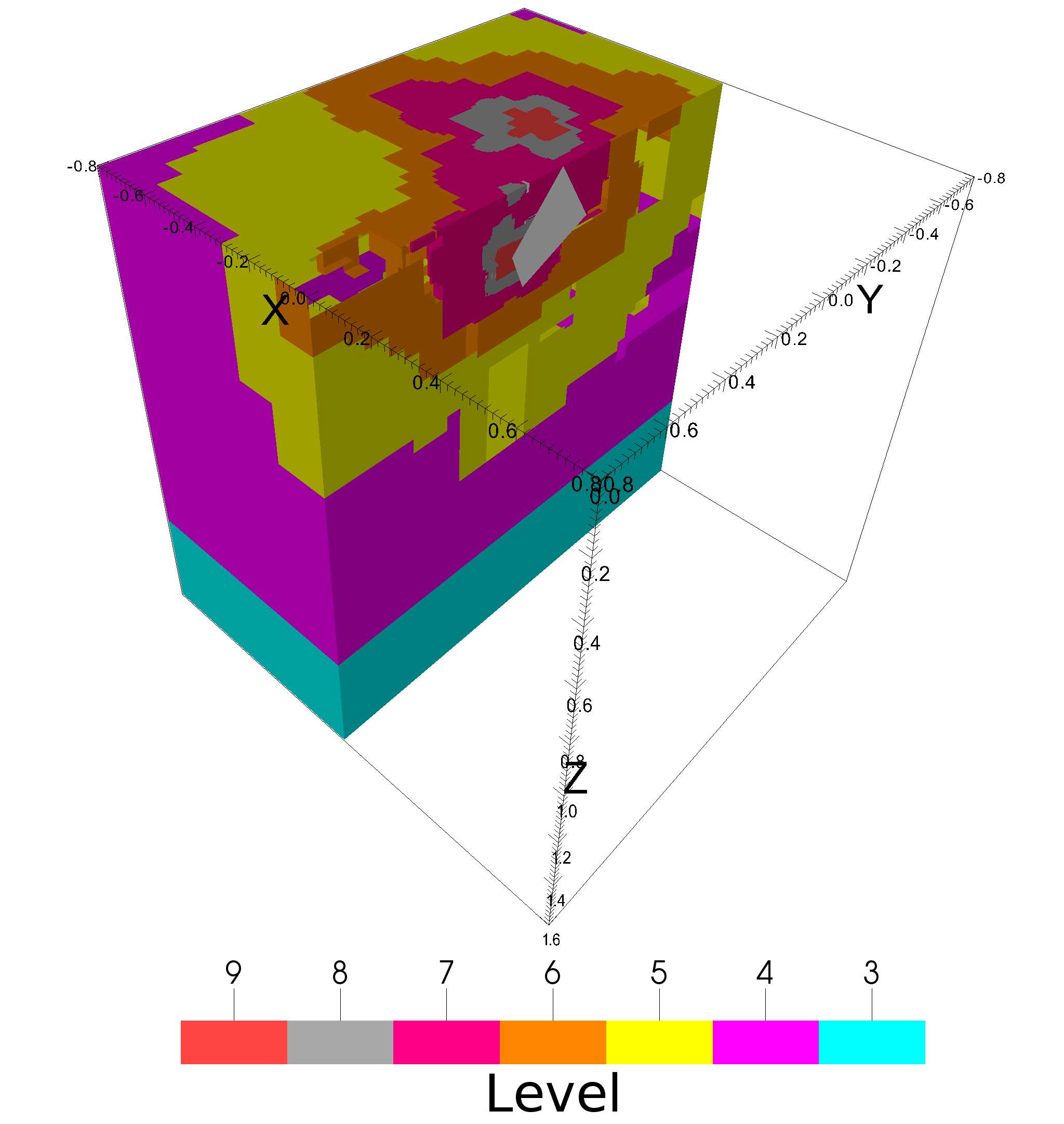}\protect\caption{A cutout of the second invariant of the scaled deviatoric stress $J_{2}=\left(\sigma_{ij}\sigma_{ji}-\sigma_{ii}\sigma_{jj}/3\right)/2$
of a computed solution and its associated adapted mesh levels for
a single fault in 3D. The equivalent resolution of the finest level
is $128\times128\times128$. The fault, indicated in grey, is inclined
about 25 degrees from vertical, has slip $s=10$, and has dimensions
$L=0.50$, $W=0.25$. The moduli are constant ($\mu=\lambda=1$).
We set the boundary conditions (normal Dirichlet and shear stress)
from Okada's analytic solution. The mesh is fully refined only at
the edges of the fault segment where the solution is singular. In
the center of the fault segment, the solution is discontinuous but
otherwise well behaved. So those center areas do not require full
refinement. \label{fig:okada3D_solution}}
\end{figure}

\begin{figure}
\begin{centering}
\includegraphics[width=0.9\columnwidth]{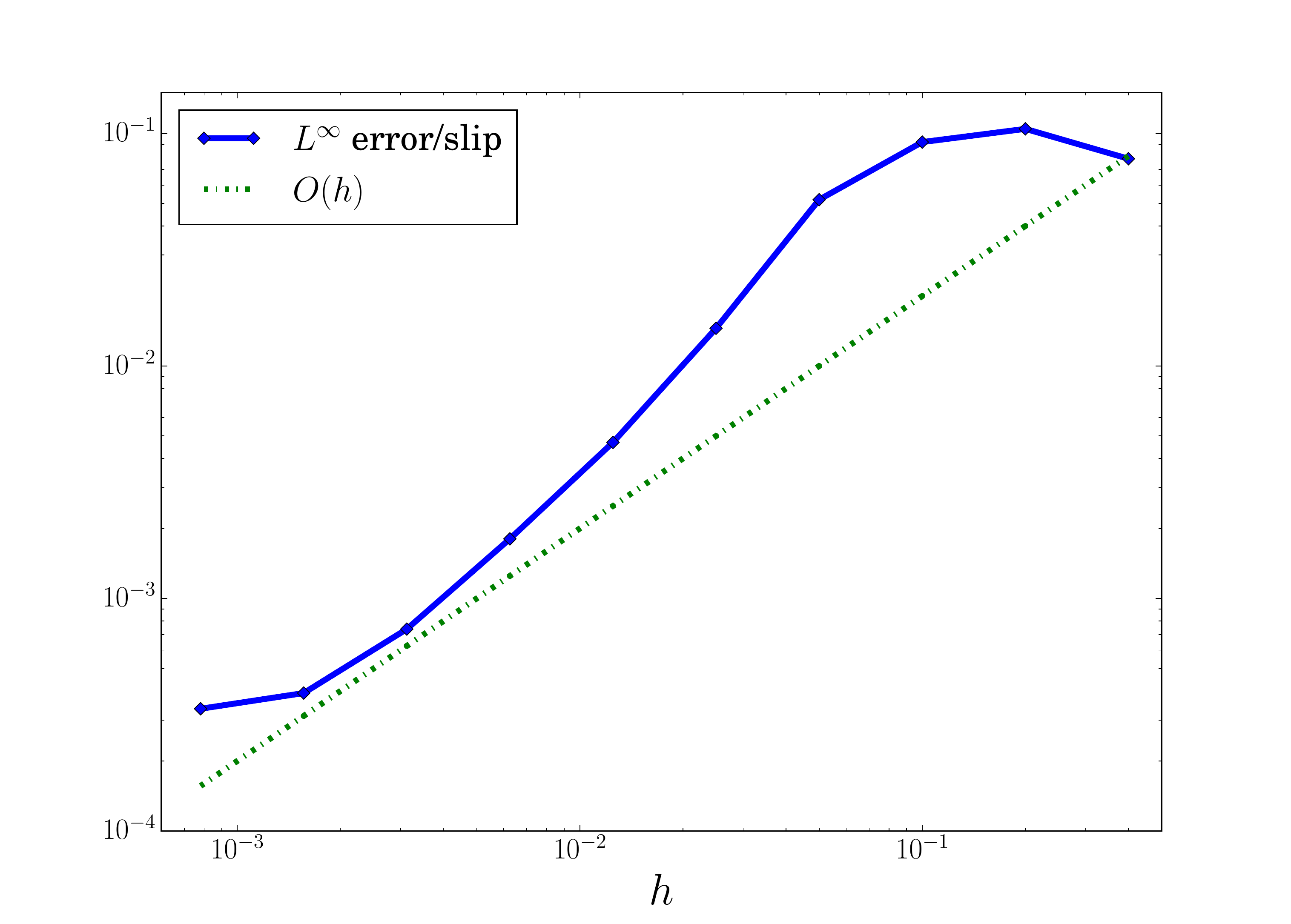} 
\par\end{centering}
\protect\caption{$L^{\infty}$ error scaled by the slip $s=10$ as a function of resolution
for the $x$ component of the displacement for a single fault in 3D.
The points within a radius of 0.1 of the side and bottom edges of
the faults are excluded. The $L^{\infty}$ error stops converging
as $O\left(h\right)$ when it becomes comparable to the adaptivity
criteria ($10^{-3}$). \label{fig:okada3D_L_inifinity}}
\end{figure}

\begin{figure}
\begin{centering}
\includegraphics[width=0.9\columnwidth]{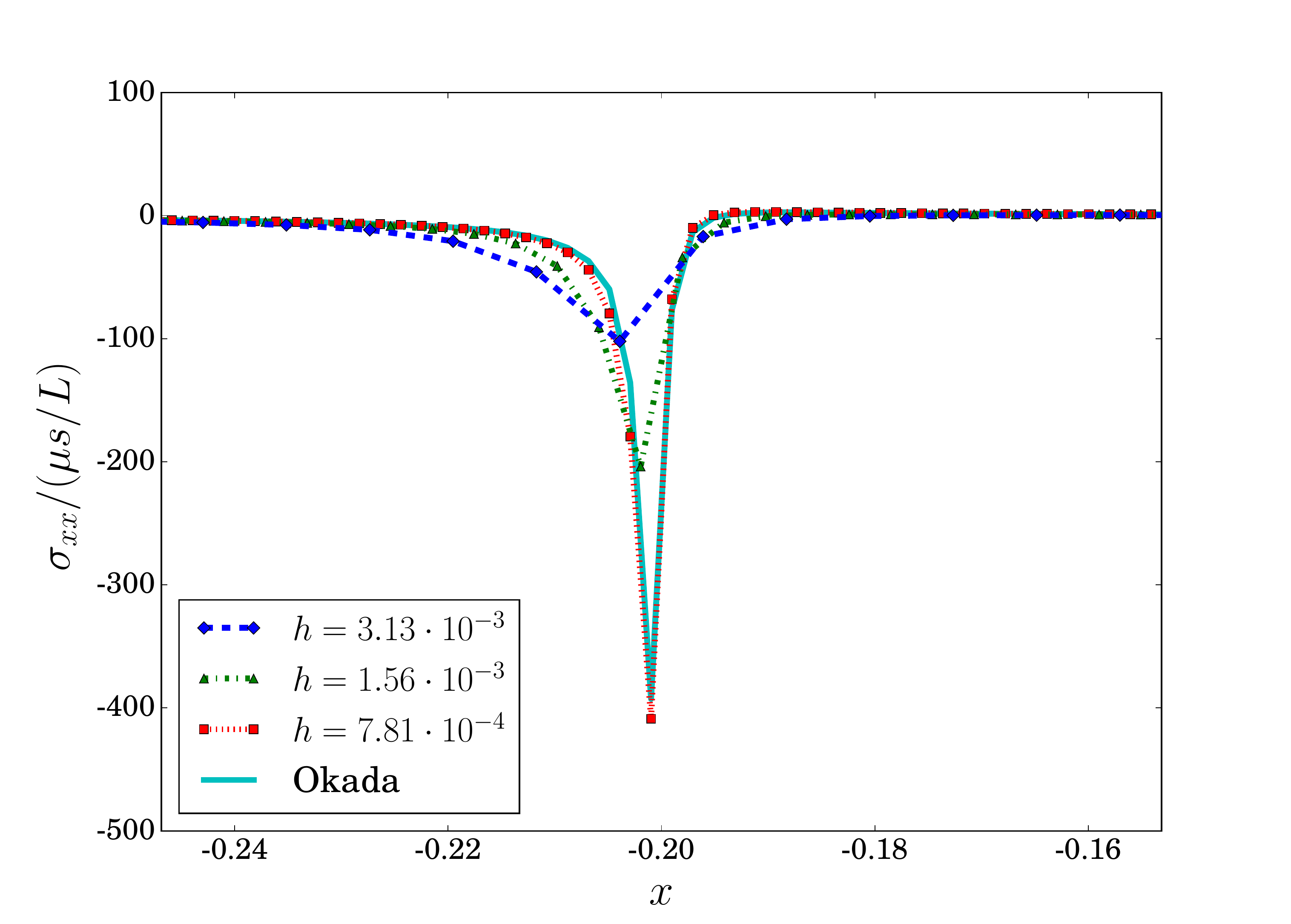} 
\par\end{centering}
\protect\caption{Numerical and analytic solutions for the scaled stress $\sigma_{xx}/\left(\mu s/L\right)$
due to a single inclined 3D fault for various resolutions. The points
are plotted along the line $y=-0.2+h/2$, $z=h/2$, passing near the
singularity in the displacement at $\left(-0.2001,-0.2001,0\right)$.
The points are offset by $h/2$ because of the staggered mesh. The
Okada solution is plotted along the same line as the finest resolution.
\label{fig:okada3D_line_section}}
\end{figure}

Okada \cite{okada85,okada92} derived an analytic expression for the
displacement due to a single fault in a homogeneous elastic half space.
Figure \ref{fig:okada3D_solution} shows a solution computed by Gamra
for an inclined, rotated fault. As the grid size gets more refined,
the mesh places points closer and closer to the singularity. This
means that the global $L^{\infty}$ error does not shrink, but rather
grows with finer resolution. To get around this, we cut holes around
the singularities and compute the $L^{\infty}$ error on that region.
Figure \ref{fig:okada3D_L_inifinity} shows the $L^{\infty}$ error
as a function of resolution. We see that the error scales as $O\left(h\right)$
up to the point where the error becomes comparable to the criteria
for adapting the mesh. Moreover, Figure \ref{fig:okada3D_line_section}
shows that, for a line crossing near the singularity in the displacement,
the stress is well behaved.

We have also run tests where we replace one of the normal Dirichlet
conditions (Section \ref{subsec:Dirichlet_BC}) with a normal stress
boundary condition (Section \ref{subsec:Stress_BC}) set using the
exact Okada stress. Similarly, we ran tests which replaced one of
the shear stress conditions (Section \ref{subsec:Stress_BC}) with
a tangential Dirichlet condition (Section \ref{subsec:Dirichlet_BC}).
All of these tests converge in a similar manner.

\begin{figure}
\begin{centering}
\includegraphics[width=0.9\columnwidth]{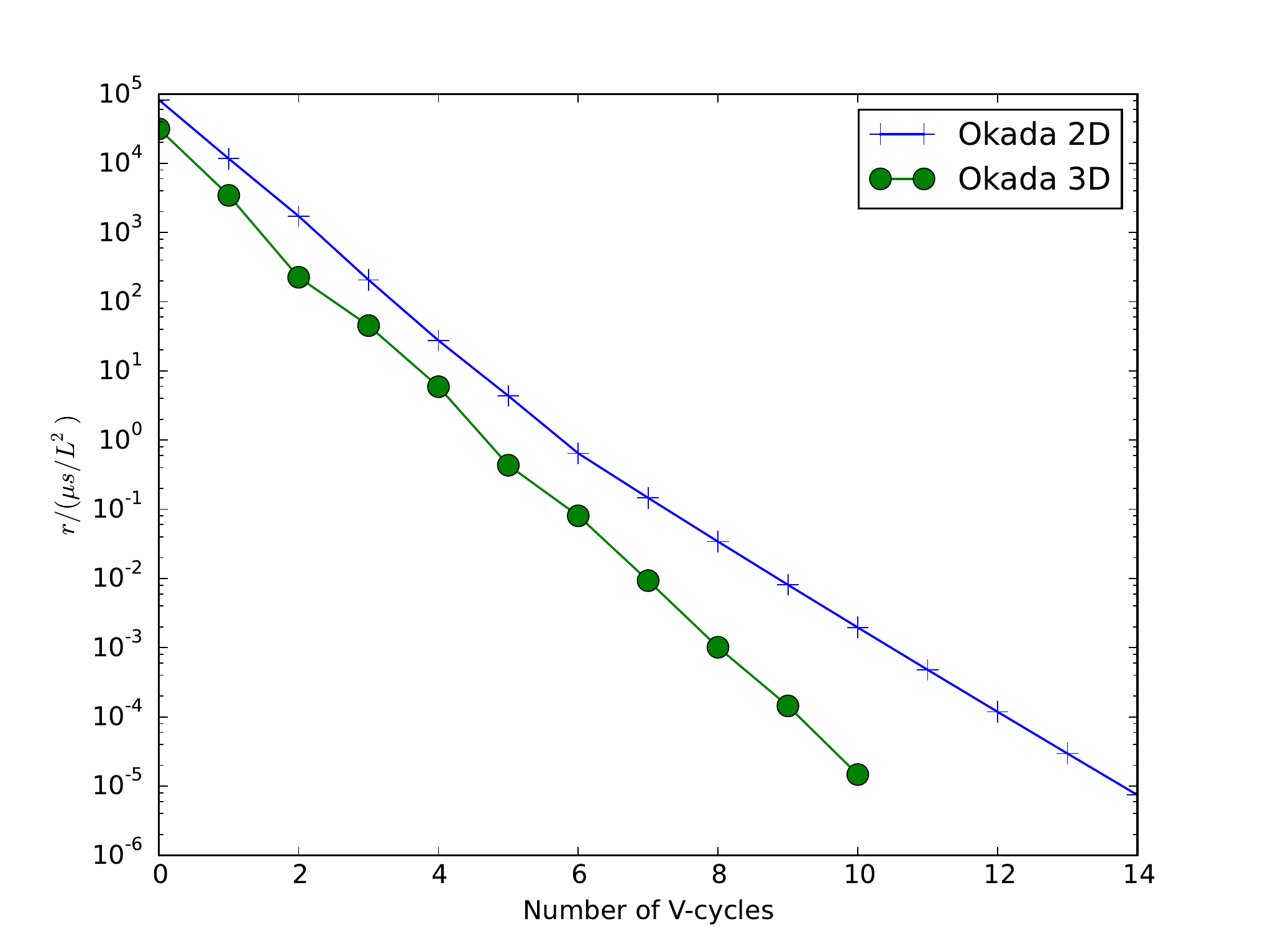} 
\par\end{centering}
\protect\caption{The scaled residual $r/\left(\mu s/L^{2}\right)$ versus the number
of multigrid V-cycles for the 2D and 3D Okada solutions.\label{fig:Okada_residuals}}
\end{figure}

Figure \ref{fig:Okada_residuals} shows the residual versus the number
of multigrid V-cycles for 2D and 3D. In spite of the singularity at
the fault tips, the solvers perform well, with the per-iteration reduction
of the residual tending asymptotically to about 0.25 in 2D and 0.12
in 3D. The 3D solver uses 4 rather than 2 sweeps per multigrid level,
so the absolute reduction in the residual is larger.

This gives us some confidence that all of the moving parts involved
in computing the solution: smoothing (Section \ref{subsec:Gauss-Seidel-Relaxation}),
boundary conditions (Section \ref{subsec:Boundary-Conditions}), multigrid
(Section \ref{subsec:Multigrid}), and adaptivity (Section \ref{subsec:Adaptive-Solution})
are correct and implemented correctly.

\section{1992 Mw 7.3 Landers earthquake}

\subsection{Setup}

\begin{figure}
\centering{}\includegraphics[width=0.45\columnwidth]{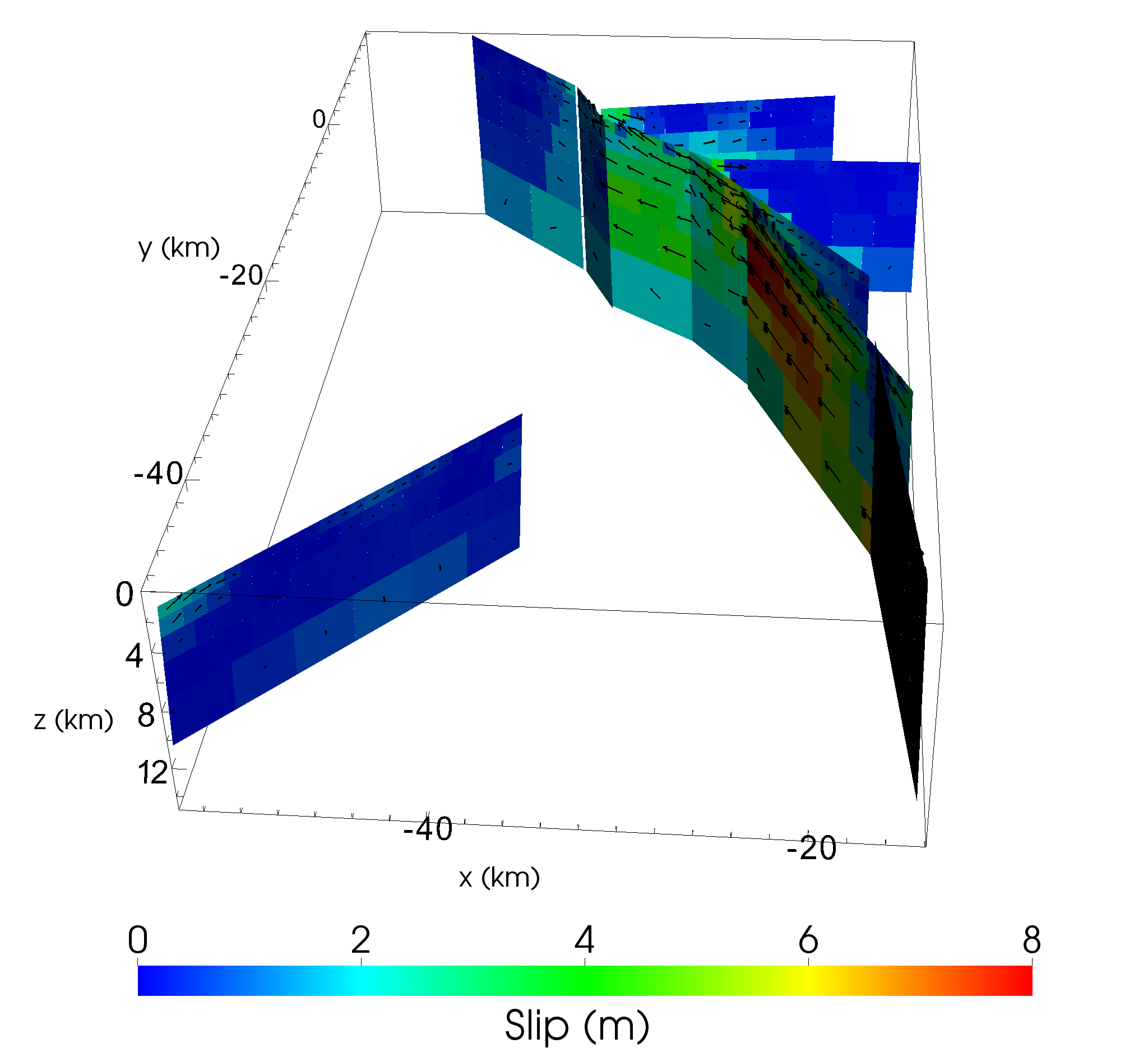}\includegraphics[width=0.45\columnwidth]{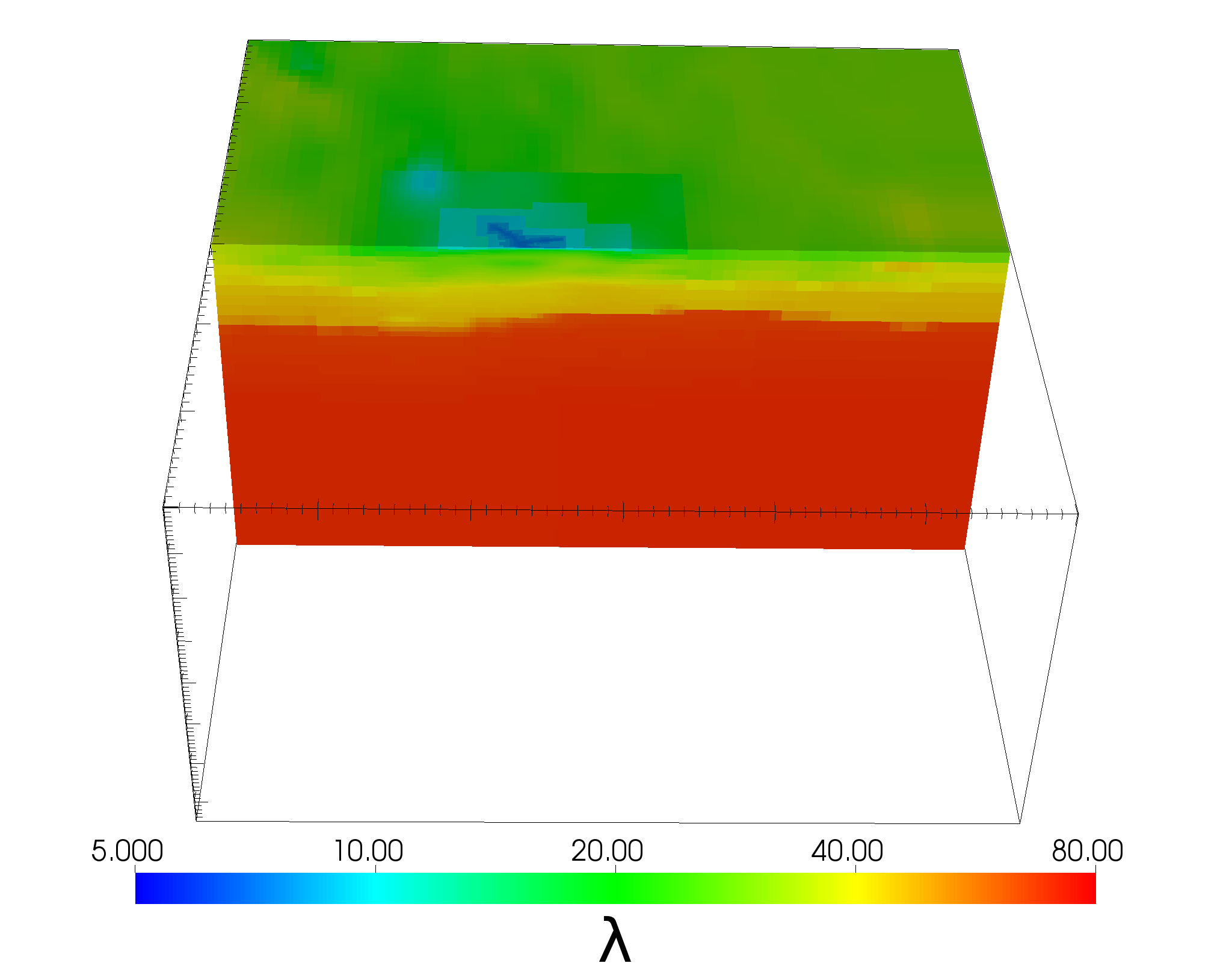}\caption{\label{fig:Landers-faults}Fault geometry, slip and $\lambda$ for
the 1992 Mw 7.3 Landers model \cite{fialko04b}. }
\end{figure}

We construct a realistic model of the 1992 Mw 7.3 Landers earthquake
using the slip model from Fialko \cite{fialko04b} and the material
model from the Southern California Earthquake Center Community Velocity
Model - Harvard (CVM-H) \cite{tape+09}. The slip model consists of
426 individual fault segments (Figure \ref{fig:Landers-faults}).
Figure \ref{fig:Landers-faults} also shows the variation of Lame's
first parameter, $\lambda$. The second Lame parameter, $\mu$, has
similar structure.

The boundaries are about 100-200 km away from the faults. The boundary
conditions on the sides and bottom are free slip: zero shear stress
(Section \ref{subsec:Stress_BC}) and zero normal displacement (Section
\ref{subsec:Dirichlet_BC}). The boundary condition on the top is
free surface: zero shear and normal stress (Section \ref{subsec:Stress_BC}).
Since these boundary conditions are imperfect, the error due to the
boundaries is about the size of the displacement at the boundary:
1 cm. Getting the error down to the current limits of GPS technology
(about 0.5 mm \cite{hill+09,langbein08a,williams+04}), would require
moving the boundaries so far away such that other effects not accounted
for (e.g. topography, curvature of the earth) would become significant.

During a multigrid V-cycle, we used 4 pre- and post- sweeps. On the
coarsest level, we smoothed 32 times to get an approximate solution.
We set the refinement criteria $\epsilon$ (Section \ref{subsec:Adaptive-Solution})
to our estimate of the boundary error: 1 cm. We continue multigrid
V-cycles until the $L^{\infty}$ norm of the residual (Eq.~\ref{eq:residual})
is less than $10^{-3}\text{\,\ m}\,\text{GPa}\,\text{km}^{-2}$. From
equation \ref{eq:minimum_dx}, this implies a minimum resolution of
$\sqrt{7\cdot0.01/10^{-3}}=8.37\text{\ km}$, which in this case is
satisfied when the refinement level is at least 3. The mesh is globally
refined to level 3, so the error is always dominated by the discretization.

\subsection{Results}

\begin{figure}
\begin{centering}
\includegraphics[width=0.9\columnwidth]{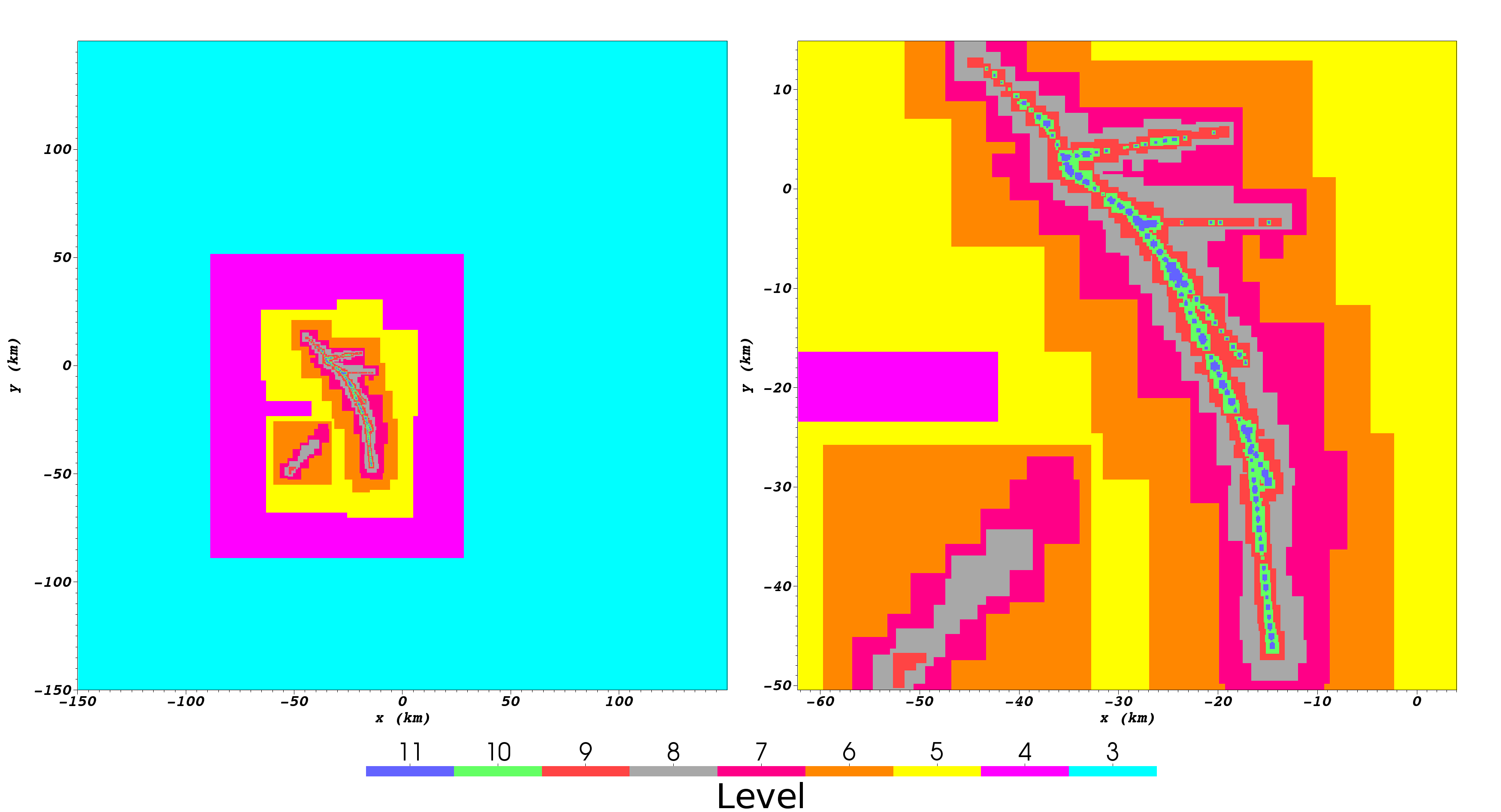} 
\par\end{centering}
\protect\caption{\label{fig:Landers-levels}AMR level hierarchy for the 1992 Landers
model at the surface ($z=0$) of the model (left) and in a zoomed
in region around the faults (right). Level 3, the coarsest level shown,
has $64\times64\times32$ elements with a resolution of 4700 meters.
Level 11 has a resolution of 18 meters.}
\end{figure}

\begin{figure}
\begin{centering}
\includegraphics[width=0.9\columnwidth]{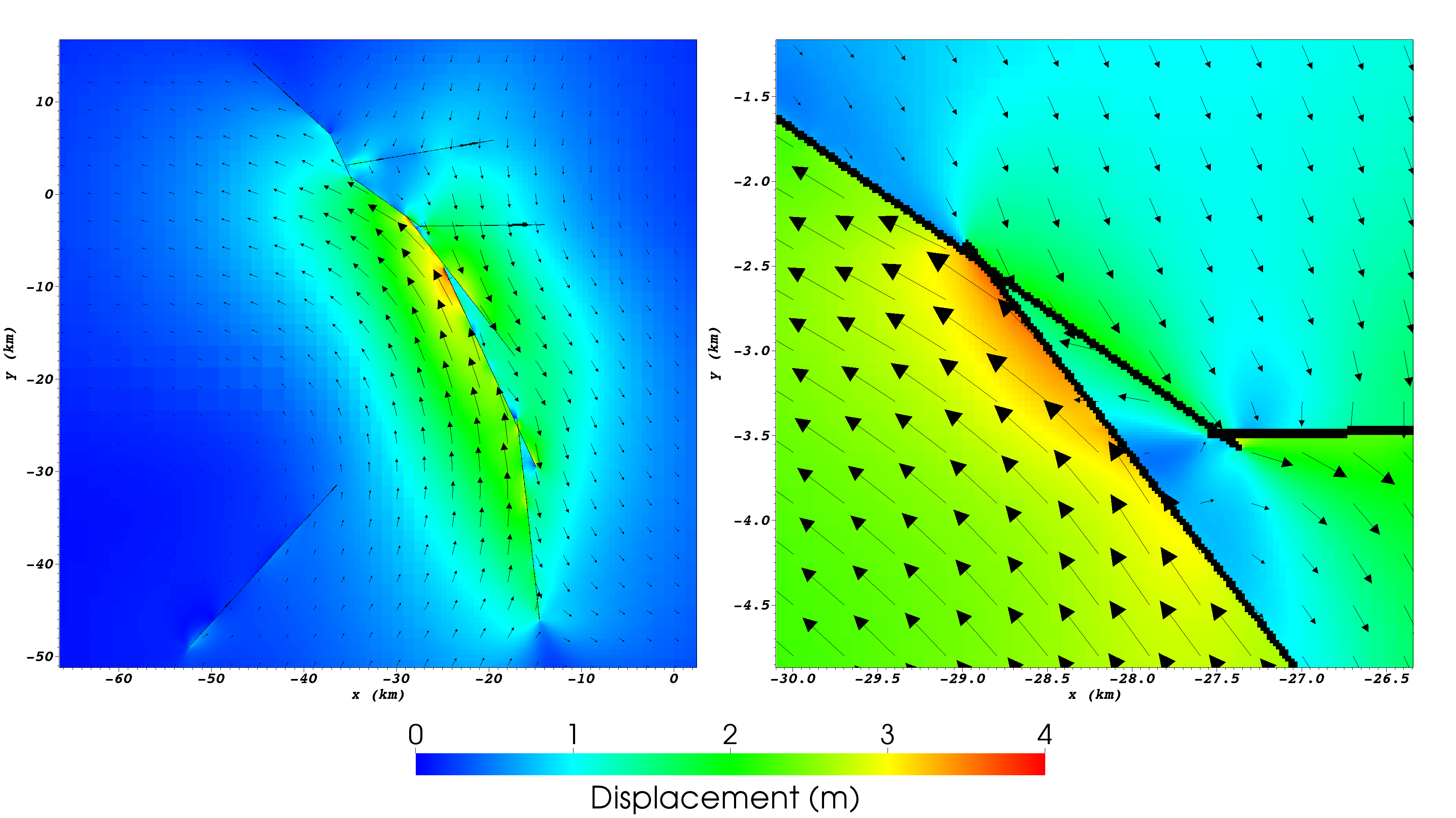} 
\par\end{centering}
\protect\caption{\label{fig:Landers-displacement}Zoomed in views of the computed surface
displacement for the 1992 Landers model. The black lines indicate
fault segments.}
\end{figure}

Gamra automatically generated the highly adapted mesh in Figure \ref{fig:Landers-levels}.
This mesh has $8.1\times10^{7}$ elements, while an equivalent non-adaptive
mesh would require $2.2\times10^{12}$ elements. The computed solution
in Figure \ref{fig:Landers-displacement} highlights the discontinuous
nature of the solutions. We expect the error to be concentrated close
to the faults, as in Figure \ref{fig:okada3D_L_inifinity}. So even
though the error may be larger near the faults, this would not translate
to a large offset error farther from the faults. With that in mind,
we expect that the error in displacement in the regions covered by
levels 3-10 to be about 1 cm, or about 0.125\% of the maximum displacement.
Otherwise, the automatic refinement criteria would have marked those
regions for refinement.

\subsection{Performance}

\begin{figure}
\begin{centering}
\includegraphics[width=0.9\columnwidth]{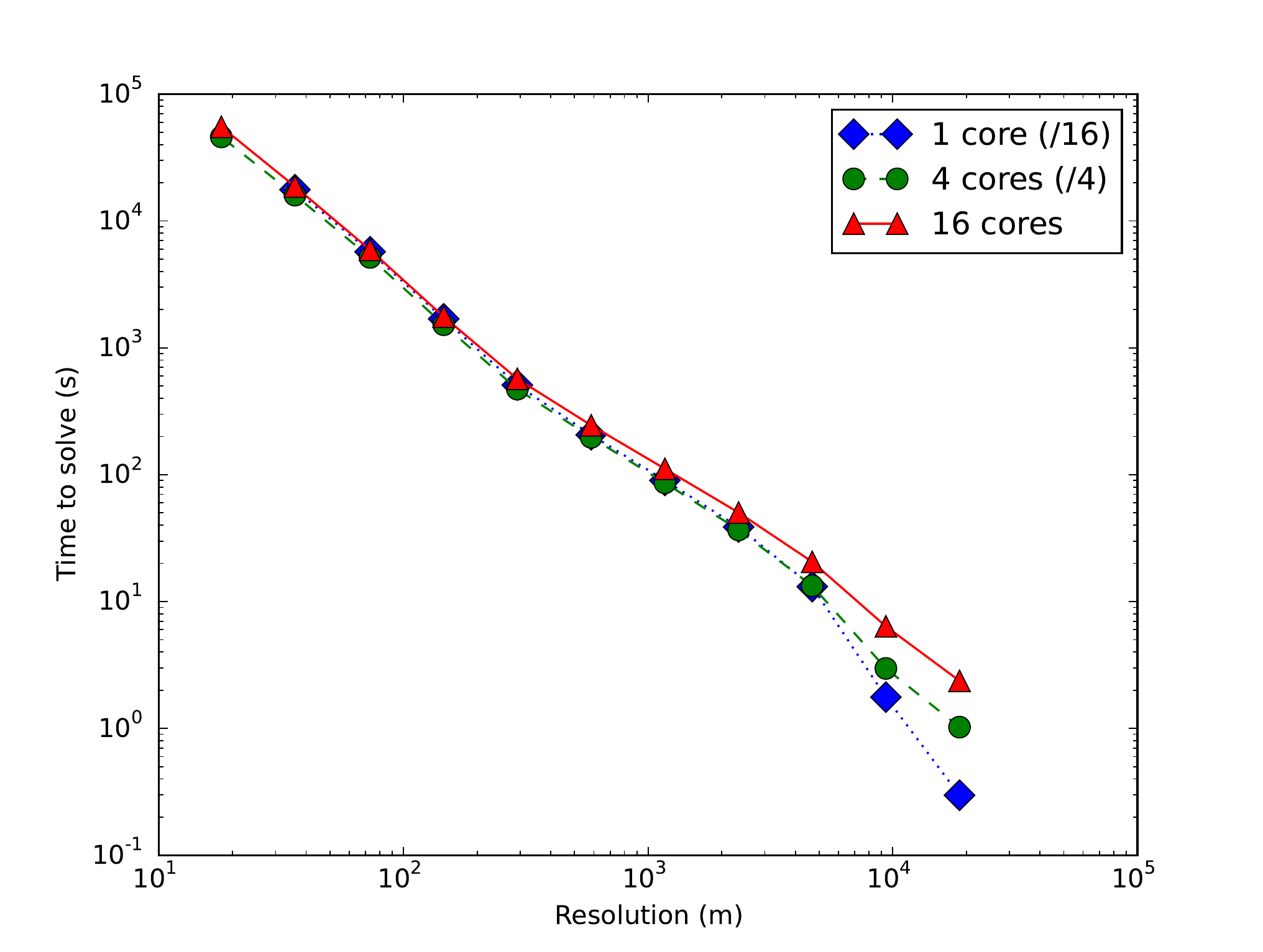} 
\par\end{centering}
\protect\caption{\label{fig:Landers_timing}Scaled performance for the Landers model.}
\end{figure}

We computed this Landers earthquake solution on a Xen virtual machine
running in a Dell R720 with 16 physical cores (Intel Xeon CPU E5-2670)
and 256 GB of RAM using OpenMPI 1.8.8 and gcc 4.7.3. Figure \ref{fig:Landers_timing}
shows the time to solve as a function of resolution and number of
cores. Altogether, the scaling is quite good at finer resolutions
on this shared memory architecture.

Although it is difficult to see in the plot, we see superlinear scaling
from 1 to 4 cores for finer resolution. This superlinear scaling does
not persist for higher core counts. This is probably a quirk due to
running inside a virtual machine. On different hardware without a
virtual machine (8 physical core Intel Xeon CPU E5620), we do not
see superlinear scaling.

We can roughly fit the relation between time and grid spacing on the
plot with a power law $t\propto h^{-1.85}.$ This is significantly
better than a solver on a fixed three-dimensional grid. Even an optimal
multigrid solver would scale as $t\propto h^{-3}$.

\section{Conclusion}

Elastic deformation due to the displacement of faults can be modeled
efficiently with parallel multigrid methods using adaptive meshes
and embedded interfaces. The multigrid efficiency is commensurate
with what is expected for the simpler Poisson's equation multigrid
solvers~\cite{trottenberg+00}, in spite of the added complexity
brought by internal dislocations and mixed boundary conditions. The
computational efficiency is improved by the mesh adaptivity, which
reduces the number of nodes by orders of magnitudes compared with
uniform meshes. A key advantage of the proposed method is the ability
to simulate complex fault geometries without manual and labor-intensive
meshing. Even in these complex models, we experienced no problems
due to instabilities in the solver or excess sensitivity of the final
solution to small changes in the input.

In addition, the method offers high precision in the near field of
faults, even capturing the stress singularity asymptotically (Figure
\ref{fig:okada3D_line_section}). This is important for evaluating
stress and other dynamic variables. All of these features make the
proposed approach optimal for generating stress and displacements
kernels for inversions for fault slip \cite{barbot+13}, investigation
of the surrounding elastic structure \cite{barbot+09b,cochran+09},
and building stress and displacement kernels for simulations using
the boundary-integral method.

This study presents an important building block of earthquake cycle
simulations. A future major undertaking will be to incorporate rupture
dynamics and quasi-static off-fault deformation. Fault dynamics will
require modeling the propagation of seismic waves. The mesh adaptivity
may then be exploited to implement spatially variable adaptive time
steps \cite{meglicki+07}. Quasi-static time-dependent problems with
off-fault plasticity and visco-elastic or poro-elastic deformation
may be treated with the same elliptic solver using equivalent body
forces (per unit time), requiring only more book-keeping to handle
explicit time steps. Many other effects may be incorporated to enable
even more realistic models of earthquakes and Earth deformation, such
as a spherical geometry for global-scale models and topography to
improve calculation of local stress.

\section{Acknowledgements}

We thank two anonymous reviewers for their comments that improved
this manuscript. This research used resources of the Argonne Leadership
Computing Facility, which is a DOE Office of Science User Facility
supported under Contract DE-AC02-06CH11357. This research was supported
by the Gordon and Betty Moore Foundation, NSF Award EAR-0949446, the
Tectonics Observatory, the National Research Foundation of Singapore
under the NRF Fellowship scheme (National Research Fellow Award No.
NRF-NRFF2013-04) and by the Earth Observatory of Singapore and the
National Research Foundation and the Singapore Ministry of Education
under the Research Centres of Excellence initiative. This is EOS publication
113.

\appendix

\section{Adaptive Multigrid\label{sec:Adaptive-Multigrid}}

For completeness, we detail the exact adaptive multigrid algorithm
we use. This is mostly a restatement of Section 4 of Martin \& Cartwright
\cite{martin&cartwright96}.

First, we define a Gauss-Seidel operator $\text{GS}\left(\vec{v},\vec{f},N\right)$,
where $\vec{v}$ is an initial guess, $\vec{f}$ is the forcing term,
and $N$ is the number of times to apply the smoother. The output
of $\text{GS}\left(\vec{v},\vec{f},N\right)$ is a correction

\begin{equation}
\delta\vec{v}=\text{GS}\left(\vec{v},\vec{f},N\right)\,.\label{eq:GS_operator}
\end{equation}
For $N=1$, Equation \ref{eq:v_update} implies

\[
\text{GS}\left(\vec{v},\vec{f},1\right)_{i}=\frac{-r_{i}\left(\vec{v},\vec{f}\right)}{dr_{i}/dv_{i}}\,,
\]
where $r_{i}\left(\vec{v},\vec{f}\right)$ is defined by Equation
\ref{eq:residual}. Next we define a recursive multigrid V-cycle relaxation
routine $\text{MGRelax}\left(l,\vec{d}\right)$, where $l$ is the
current level and $\vec{d}$ is the defect. The outline of the routine
is as follows. 
\begin{enumerate}
\item If $l=0$ (the coarsest level)

\begin{enumerate}
\item Using an initial guess of 0, compute a correction by applying the
smoother $N_{\text{coarse}}$ times 
\[
\delta\vec{v}=\text{GS}\left(\vec{0},\vec{d},N_{\text{coarse}}\right)
\]
or until the $L^{\infty}$ norm of the residual $r_{i}(\delta\vec{v},\vec{d})$
is less than $\epsilon_{\text{coarse}}$. 
\end{enumerate}
\item If $l>0$

\begin{enumerate}
\item Using an initial guess of 0, compute a correction $\delta\vec{v}$
by applying the smoother $N_{\text{pre}}$ times 
\[
\delta\vec{v}=\text{GS}\left(\vec{0},\vec{d},N_{\text{pre}}\right)\,.
\]
\item Compute $r_{i}\left(\delta\vec{v},\vec{d}\right)$, the residual on
just the fine grid (Section \ref{subsec:Gauss-Seidel-Relaxation}). 
\item Coarsen $r_{i}$ to make $R_{i}$ (Section \ref{subsec:Coarsening})
\[
R_{i}=\text{Coarsen}\left(r_{i}\right)\,.
\]
\item Recursively call MGRelax to get the coarse grid correction 
\[
\delta\vec{V}=\text{MGRelax}\left(l-1,R_{i}\right)\,.
\]
\item Refine the correction $\delta\vec{V}$ to the fine level (Section
\ref{subsec:Refinement}) and add it to the fine correction $\delta\vec{v}$
\[
\delta\vec{v}=\delta\vec{v}+\text{Refine}\left(\delta\vec{V}\right)\,.
\]
\item Apply the smoother $N_{\text{post}}$ times to get a final correction
\[
\delta\vec{v}=\text{GS}\left(\delta\vec{v},\vec{d},N_{\text{post}}\right)\,,
\]
\end{enumerate}
\item Return $\delta\vec{v}$. 
\end{enumerate}
Given these functions, the driver routine is short. 
\begin{enumerate}
\item Compute a composite residual $r_{i}$ (Equation \ref{eq:residual}).
This includes applying all physical (Section \ref{subsec:Boundary-Conditions})
and coarse-fine (Section \ref{subsec:Coarse-Fine-boundaries}) boundary
conditions. 
\item While the $L^{\infty}$ norm of the residual is less than the stopping
tolerance $\epsilon_{\text{stopping}}$

\begin{enumerate}
\item Compute $\delta\vec{v}=\text{MGRelax}\left(l_{\max},r_{i}\right)$. 
\item Add in the correction 
\[
\vec{v}=\vec{v}+\delta\vec{v}
\]
\item Recompute the composite residual $r_{i}$. 
\end{enumerate}
\end{enumerate}
In pseudo-code, $\text{MGRelax}$ is
\begin{lyxcode}
Procedure~$\text{MGRelax}\left(l,\vec{d}\right)$:

~~if~$l>0$:

~~~~$\delta\vec{v}=0$

~~~~for~(i~=0;~i<$N_{\text{coarse}}$;~++i)

~~~~~~$\delta\vec{v}=\delta\vec{v}+\text{GS}\left(\delta\vec{v},\vec{d},1\right)$

~~~~~~if~$\left(L^{\infty}\left(r_{i}\left(\delta\vec{v},\vec{d}\right)\right)<\epsilon_{\text{coarse}}\right)$:

~~~~~~~~break

~~else:

~~~~$\delta\vec{v}=\text{GS}\left(\vec{0},\vec{d},N_{\text{pre}}\right)$

~~~~$R_{i}=\text{Coarsen}\left(r_{i}\left(\delta\vec{v},\vec{d}\right)\right)$

~~~~$\delta\vec{V}=\text{MGRelax}\left(l-1,\vec{R}\right)$

~~~~$\delta\vec{v}=\delta\vec{v}+\text{Refine}\left(\delta\vec{V}\right)$

~~~~$\delta\vec{v}=\text{GS}\left(\delta\vec{v},\vec{d},N_{\text{post}}\right)$

~~return~$\delta\vec{v}$
\end{lyxcode}
and the driver is
\begin{lyxcode}
while~$\left(L^{\infty}\left(r_{i}\left(\vec{v},\vec{d}\right)\right)>\epsilon_{\text{stopping}}\right)$

~~$\vec{v}=\vec{v}+\text{MGRelax}\left(l_{\text{max}},\vec{r}\right)$
\end{lyxcode}
\bibliographystyle{abbrv}
\bibliography{reference}

\end{document}